\documentclass[reprint,aps,prb,twocolumn]{revtex4-1}
\usepackage{graphicx}

\usepackage[english]{babel}
\usepackage{amssymb,amsfonts,psfrag,amsmath,bm,bbm,indentfirst,subfigure,bbold}
\usepackage{color}
\usepackage[colorlinks,linkcolor=blue,urlcolor=black,citecolor=blue]{hyperref}
\usepackage{ulem}
\usepackage{natbib}

\definecolor{darkgreen}{rgb}{0,0.5,0}
\definecolor{purple}{rgb}{0.35,0,0.35}
\definecolor{orange}{rgb}{1,0.5,0}
\definecolor{darkred}{rgb}{.7,0,0}
\definecolor{darkblue}{rgb}{0.1,0.1,.6}
\definecolor{grey}{rgb}{.6,.6,.6}
\definecolor{dimgreen}{rgb}{0.2,0.6,0.1}
\definecolor{RoyalBlue}{cmyk}{0.94,0.539,0,0}
\definecolor{DGLorange}{cmyk}{.22,1,1,.2}
  
\newcommand{\Chi}{{\rm X}}
\newcommand{\bubble}{{W}}
\newcommand{\Eq}[1]{Eq.~(\ref{#1})}
\newcommand{\Eqs}[1]{Eqs.~(\ref{#1})}

\renewcommand{\emph}[1]{\textit{#1}}
\begin{document}

\title{Functional Renormalization Group Approach for Inhomogeneous
Interacting Fermi-Systems}

\author{Florian Bauer}
\author{Jan Heyder}
\author{Jan von Delft}

\affiliation{Arnold Sommerfeld Center for Theoretical Physics and Center for
NanoScience,
Ludwig-Maximilians-Universit\"at
M\"unchen, Theresienstrasse 37, D-80333 M\"unchen, Germany}

\date{January 1, 2014}

\begin{abstract}
  The functional renormalization group (fRG) approach has the property that,
  in general, the flow equation for the two-particle vertex generates
  $\mathcal{O}(N^4)$ independent variables, where $N$ is the number of
  interacting states (e.g.\ sites of a real-space discretization). In
  order to include the flow equation for the two-particle vertex one
  needs to make further approximations if $N$ becomes too large. We
  present such an approximation scheme, called the coupled-ladder
  approximation, for the special case of an onsite interaction.  Like the
  generic third-order-truncated fRG, the coupled-ladder approximation
  is exact to second order and is closely related to a simultaneous
  treatment of the random phase approximation in all channels,
  i.e. summing up parquet-type diagrams. The scheme is applied to a
  one-dimensional model describing a quantum point contact.
\end{abstract}
\maketitle

\section{Introduction}
The  calculation of properties of an inhomogeneous
interacting quantum system requires adequate care regarding a proper
description of its spatial structure: for a lattice model the
resolution of a potential landscape, without generating additional
finite size effects, typically requires an extension of $\sim\! 10^2$
sites per spatial dimension.  If, in addition, the strength of
interactions cannot be regarded as `weak', a reasonable approximation
scheme must involve detailed information about higher order
correlations. This usually demands a huge effort for modern computers,
both in memory and speed. Thus, for a system with non-trivial
  spatial structure any approximation scheme necessarily involves a
  tradeoff between computational feasibility and accuracy.

  In Ref.~\onlinecite{Bauer2013}, we introduced such a scheme, both
  reasonably fast and accurate up to intermediate interaction
  strength, within the framework of the one-particle
    irreducible version of the functional renormalization group
  (fRG).\cite{Wetterich1993,Metzner2011,Meden2002,Andergassen2004,Andergassen2006,Karrasch2008,Karrasch2006,Husemann2012,Giering2012}
  The goal of this paper is to supply a detailed description of
  this approximation scheme, called the coupled-ladder approximation
  (CLA), which is implemented within the context of generic,
  third-order-truncated fRG.  In the latter, the flow of the
  three-particle vertex is set to zero, while the flow equation of the
  two-particle vertex (which we will call ``vertex flow'' in the
  following) is fully incorporated. This vertex flow has to be
  incorporated if interactions cannot be considered small.  In
  general, this constitutes a computational challenge, since the
  vertex generated by this flow involves a large number $\mathcal{O}
  (N^4)$ of independent functions, each depending on three
    frequencies, where $N$ is the number of sites of
  the interacting region. As a result, the flow equations
    involve $\mathcal{O} (N^4N_{\rm f}^3)$ independent variables,
    where $N_{\rm f}$ is the number of discrete points per
      frequency used in the numerics.  Previous schemes
  that included the vertex flow for models with large $N$ made use of
  an additional symmetry, e.g.
  Ref.~\onlinecite{Andergassen2004,Andergassen2006} described systems
  with a weak spatial inhomogeneity (either changing
    adiabatically with position, or confined to a small region),
  which could be treated as a perturbation, so that its feedback to
  the vertex could be neglected. The resulting equations for the
  vertex were solved in the momentum basis, exploiting the
    fact that the single-particle eigenstates could approximately be
    represented by plane waves.  However, this is not possible for
  models with strong inhomogeneities. Our CLA scheme was developed to
  include the vertex flow for such models.  It extends the idea of
  Refs.~\onlinecite{Karrasch2008}~and~\onlinecite{Jakobs2010}, where
  the CLA was introduced to parametrize the frequency dependence of
  the vertex for the single-impurity Anderson model,
  i.e. $N=1$, which reduces the number of independent
    variables for that model to $\mathcal{O} (N_{\rm f})$.  We show
  that the CLA can be applied to parametrize the spatial dependence of
  the vertex for models with a purely local interaction. The number of
  independent variables that represent the spatial dependence of the
  vertex then reduces to $\mathcal{O} (N^2)$, and the
      total number of independent variables representing the vertex to
      $\mathcal{O}(N^2 N_{\rm f})$.  The CLA scheme is exact to
  second order\cite{Honerkamp2001a,Honerkamp2003} and effectively sums
  up diagrams of the Random Phase Approximation (RPA) of all three
  interaction channels.

To illustrate the capabilities of our CLA scheme, we apply it, as in
Ref.~\onlinecite{Bauer2013}, to a one-dimensional chain modeling the
lowest submode of a quantum point contact (QPC), a short constriction
that allows transport only in one dimension. Its conductance is
famously quantized\cite{Wees1988,Wharam1988,Buttiker1990} in units of
$G_Q={2e^2}/{h}$. In addition to this quantization, measured
conductance curves show a shoulder at around $0.7 G_Q$. In this regime
quantities such as electrical and thermal conductance, noise and
thermo-power have anomalous
behavior\cite{Thomas1996,Appleyard2000,Cronenwett2002}. These
phenomena are collectively known as the ``0.7-anomaly'' in QPCs.

In Ref.~\onlinecite{Bauer2013} we showed that the 0.7-anomaly is reproduced by a
one-dimensional model with a parabolic potential barrier and a short-ranged Coulomb
interaction. We presented a detailed microscopic picture that explained the physical
mechanism which causes the anomalous behavior. Its origin is a smeared van Hove
singularity in the density of states (DOS) just above the band bottom which enhances
effects of interaction causing an enhanced backscattering. We presented detailed
results for the conductance at zero temperature, obtained using fRG in the CLA. These
numerical data were in good qualitative agreement with our experimental
measurements and showed that the model reproduces the phenomenology of the
0.7-anomaly. In this paper we set forth and examine the approximation scheme in
detail. We present additional numerical data to verify the reliability of the method
for the case where it is applied to the model of a QPC. For this we present and
compare data obtained by different approximation schemes within the fRG, showing that
the phenomenology is very robust, and can even be obtained by neglecting the vertex
flow.  However including the vertex flow using the CLA reduces artifacts and gives an
insightful view on the spin susceptibility. For the latter, we finally present a
detailed quantitative error analysis.

\section{Microscopic Model}
The approximation scheme presented in this paper can be applied to any
model Hamiltonian that can be written in the following form:
\begin{align}
 H=\sum_{ij, \sigma} h_{ij}^{\sigma} d^\dagger_{i\sigma} d_{j\sigma}^{} +\sum_j
 U_j n_{j\uparrow} n_{j\downarrow} \; ,
 \label{eqModelGeneral}
\end{align}
where $h^\sigma$ is a real, symmetric matrix, $d^\dagger_{j\sigma}$
($d_{j\sigma}^{}$) creates (annihilates) an electron at site $j$ with spin $\sigma$
($=\uparrow,\downarrow$ or $+,-$, with $\bar{\sigma}\!=\!-\sigma$), and $n_{j\sigma}=
d^\dagger_{j\sigma} d_{j\sigma}^{}$ counts them (in general $j$ can represent any
quantum number, however for simplicity we refer to it as a site index throughout the
paper). In order to apply the CLA the necessary property of this Hamiltonian is a
short-ranged interaction.  In principle the approximation scheme can be set up for an
interaction with finite range (over several sites), however since the structure then
becomes very complicated we will only discuss the case of a purely local, i.e.,
on-site interaction in this paper as given by Eq.~\eqref{eqModelGeneral}.
Whereas the system can extend to infinity, it is crucial that the number of sites $N$
where $U_j$ is nonzero is finite and not too large , as discussed in
section~\ref{secNumericImplementation}. If the system is extended to infinity, the
effect of the noninteracting region can be calculated analytically using the
projection method (see appendix Sec.~\ref{secProjection} and
Refs.~\onlinecite{Taylor1972}~and~\onlinecite{Karrasch2006}).  An extension to a
Hamiltonian that is complex Hermitian and non-diagonal in spin space, needed, e.g.,
to include spin-orbit effects, is straightforward.  In contrast applying the scheme
to spin less models, for which the interaction term has to be non-local to respect
Pauli's exclusion principle, is more complicated.

\section{fRG flow equations}\label{secfRG}

In this section we describe the functional Renormalization Group (fRG) approach that
we have employed to treat a translationally nonuniform Fermi system with onsite
interactions, such as described by Eq.~\eqref{eqModelGeneral}. We use the one-particle
irreducible (1PI) version of the fRG\cite{Wetterich1993,Morris1994}.  Its key idea is
to approximately sum up a perturbative expansion, in our case in the interaction, by
setting up and numerically solving a set of coupled ordinary differential equations
(ODEs),
the \textit{flow equations}, for the system's 1PI $n$-particle vertex functions,
$\gamma_n$.  This is typically done in such a way that the effects of higher-energy
modes, lying above a flowing infrared cutoff parameter $\Lambda$, are incorporated
before those of lower-energy modes lying below $\Lambda$.  This yields a systematic
way of summing up parquet-type diagrams for the two-particle vertex and for
calculating the self-energy.  $\Lambda$ serves as \textit{flow parameter} that
controls the RG flow of the $\Lambda$-dependent vertex functions $\gamma_n^\Lambda$
from an initial cutoff $\Lambda_i$, at which all vertex functions are known and
simple, to a final cutoff $\Lambda_f$, at which the full theory is recovered.

This idea is implemented by replacing, in the generating functional for the
vertex functions $\gamma_n$, the bare propagator
$\mathcal{G}_0$ by a modified propagator $\mathcal{G}^\Lambda_0$,
\begin{equation}
  \mathcal{G}_0 \rightarrow \mathcal{G}_0^{\Lambda} \; , 
  \quad \textrm{with} \quad
  \mathcal{G}_0^{\Lambda_i}=0, \quad \mathcal{G}_0^{\Lambda_f}=\mathcal{G}_0 \; ,
\end{equation} 
constructed such that $\mathcal{G}_0^\Lambda$ is strongly suppressed for frequencies
below $\Lambda$. The $\Lambda$-dependence of the resulting vertex functions
$\gamma_n^\Lambda$ is governed by an infinite hierarchy of coupled ODEs, the
\textit{RG flow equations}, of the form \begin{align} \frac{d}{d \Lambda}
\gamma^\Lambda_n = \mathcal{F} \left(
\Lambda,\mathcal{G}^{\Lambda}_0,\gamma^\Lambda_1, \dots ,\gamma^\Lambda_{n+1} \right)
\; ,  \label{eq:general-flow-equations} \end{align} where $\gamma_1=-\Sigma$ is the
self-energy and $\gamma_2$ the two particle vertex. At the beginning of the RG flow,
the vertex functions are initialized to their bare values, \begin{eqnarray}
\gamma^{\Lambda_i}_2=v \quad \gamma^{\Lambda_i}_n=0 \quad (n \neq 2) ,
\end{eqnarray} while their fully dressed values, corresponding to the full theory,
are recovered upon integrating \Eqs{eq:general-flow-equations} from $\Lambda_i$ to
$\Lambda_f$.

The infinite hierarchy of ODEs (\ref{eq:general-flow-equations}) is exact, but in
most cases not solvable. In the generic, third-order-truncated fRG
all $n$-particle vertex functions with $n \geq 3$ are neglected
\begin{equation}
\label{eqfRGtrunc}
 \frac{d}{d \Lambda} \gamma_n = 0 \qquad (n \geq 3) \,,
\end{equation}
and the resulting flow equations for $\gamma_1^\Lambda$ and
$\gamma_2^\Lambda$ are integrated numerically. Due to this truncation,
fRG is in essence an ``RG-enhanced'' perturbation expansion in the
interaction, that will break down if $U$ becomes too large. In fact,
the flow equations can be derived by a purely diagrammatic procedure,
without resorting to a generating functional, as explained in
Ref.~\onlinecite{Jakobs2007}. The diagrammatic structure is such that
the flow of the self-energy and three different parquet channels
(i.e.\ three coupled RPA-like series of diagrams) are treated
simultaneously, feeding into each other during the flow (as discussed
in more detail below). This moderates competing instabilities
  in an unbiased way. We also mention that this approach has been
  found to be particularly useful to treat models where infrared
  divergences play a role\cite{Metzner2011} (though 
the latter do not arise for the present model).

The following statements in this chapter hold for most, however not for every
flow parameter. For that reason we explicitly define the $\Lambda$-dependence at this
point. If a different fRG scheme is used, one should carefully check all relations.
The general idea should be applicable for all fRG schemes.
We use fRG in the Matsubara formalism. In the following frequencies with subscripts
$n$, $n'$, $n_1$,
etc., are defined to be purely imaginary:
\begin{equation}
 \omega_n=iT\pi(2n+1) .
\end{equation}
We introduce $\Lambda$ as an infrared cutoff in the bare Matsubara propagator,
\begin{equation}
\label{eqfRGscheme}
\mathcal{G}_0^{\Lambda} (\omega_n)
 =\Theta_T (| \omega_n | - \Lambda ) \mathcal{G}_0 (\omega_n) \; ,
\; \Lambda_i = \infty, \; \Lambda_f = 0 \; ,
\end{equation}
where $\Theta_T$ is a step function that is broadened on the scale of the temperature
$T$.

For a derivation of the fRG flow equations, see, e.g.,
Refs.~\onlinecite{Metzner2011,Andergassen2004}; very detailed discussions are
given e.g. in Refs.~\onlinecite{Karrasch2006,Bauer2008}, for a diagrammatic
derivation see Ref.~\onlinecite{Jakobs2007}. The flow equation for the self-energy
reads as:
\begin{align}
\label{eqgamma1DGL}
\frac{\textrm{d}}{\textrm{d} \Lambda} \gamma_1^\Lambda
( \color{DGLorange} q_1' \color{black}, \color{RoyalBlue} q^{~}_1
 \color{black}) =
 T \sum_{q_2',q_2^{~}}
 \mathcal{S}_{q^{~}_2,q'_2}^\Lambda
 \gamma_{2}^\Lambda
(q_2' ,\color{DGLorange} q_1' \color{black}; q^{~}_2,
\color{RoyalBlue} q^{~}_1  \color{black}) \;  ,
\end{align}
where $q_1,q_2$ etc.\ label the quantum number and the
fermionic Matsubara frequency.
Here $\mathcal{S}^\Lambda$ is defined in terms of the
  scale-dependent full propagator $\mathcal{G}^{\Lambda}$,
\begin{subequations}
\label{eqfRGgreen}
\begin{align}
& \mathcal{S}^\Lambda = \mathcal{G}^\Lambda
\partial_\Lambda \left[ \mathcal{G}^{\Lambda}_0 \right]^{-1} \mathcal{G}^\Lambda
\; , \\
\label{eqFullPropergator}
& \mathcal{G}^\Lambda =
\Big[ \left[ \mathcal{G}^{\Lambda}_0 \right]^{-1} - \Sigma^{\Lambda}
\Big]^{-1}
\; .
\end{align}
\end{subequations}
For later convenience we divide the two particle vertex  $\gamma_2 $ in four parts:
\begin{equation}
\label{eq:Ansatz-vertex-four-parts}
\gamma_2^\Lambda  = v + \gamma_p^\Lambda + \gamma_x^\Lambda + \gamma_d^\Lambda \, .
\end{equation}
where $v$ is the bare vertex and $\gamma_p^\Lambda$, $\gamma_x^\Lambda$, and
$\gamma_d^\Lambda$ are
called the particle-particle channel ($P$), and the exchange ($X$) and direct ($D$)
contributions to the particle-hole channel, respectively.
They are defined via their flow-equations with $\gamma_y^{\Lambda_i}=0$:
\begin{align}
\phantom{.} \hspace{-0.7cm} \frac{\textrm{d}}{\textrm{d}
\Lambda}  \gamma_2^\Lambda
&  = \frac{\textrm{d}}{\textrm{d} \Lambda} ( \gamma_p^\Lambda
+
\gamma_x^\Lambda
+
\gamma_d^\Lambda ) \; , 
\label{eq:diagrammatic-flow-equations}
\end{align}
Explicitly, these flow equations have the following forms:
\begin{widetext}
\begin{subequations}
\label{eqgamma2DGL}
\begin{align}
\frac{\textrm{d}}{\textrm{d} \Lambda}  \gamma_p^\Lambda
(\color{DGLorange} q_1',q_2' \color{black} ;
\color{RoyalBlue} q^{~}_1,q^{~}_2 \color{black}) & =
\phantom{-}
T \sum_{q_3', q^{~}_3, q_4', q^{~}_4}
\hspace{-3mm}\gamma_2^\Lambda
(\color{DGLorange} q_1',q_2' \color{black} ;q^{~}_3,q^{~}_4)
\mathcal{S}^\Lambda_{q^{~}_3, q_3'}
\mathcal{G}^\Lambda_{q^{~}_4,q_4'}
\gamma_2^\Lambda (q_3', q_4';
\color{RoyalBlue} q^{~}_1,q^{~}_2 \color{black})  ,
\label{eqgamma2DGL_P}
\\
\frac{\textrm{d}}{\textrm{d} \Lambda}  \gamma_x^\Lambda
(\color{DGLorange} q_1',q_2' \color{black} ;
\color{RoyalBlue} q^{~}_1,q^{~}_2 \color{black}) & =
\phantom{-}
T \sum_{q_3', q^{~}_3, q_4', q^{~}_4}
\gamma_2^\Lambda
(\color{DGLorange} q_1' \color{black} ,q_4';
q^{~}_3, \color{RoyalBlue} q^{~}_2 \color{black})
\Bigl[\mathcal{S}^\Lambda_{q^{~}_3,q_3'}
\mathcal{G}^\Lambda_{q^{~}_4, q_4'}  
+\mathcal{G}^\Lambda_{q^{~}_3,q_3'}
\mathcal{S}^\Lambda_{q^{~}_4, q_4'}  \Bigr]
\gamma_2^\Lambda (q_3', \color{DGLorange} q_2' \color{black} ;
\color{RoyalBlue} q^{~}_1 \color{black} , q^{~}_4)\; ,
\label{eqgamma2DGL_X}
\\
\frac{\textrm{d}}{\textrm{d} \Lambda}  \gamma_d^\Lambda
(\color{DGLorange} q_1',q_2' \color{black} ;
\color{RoyalBlue} q^{~}_1,q^{~}_2 \color{black}) & =
-
T \sum_{q_3', q^{~}_3, q_4', q^{~}_4}
\gamma_2^\Lambda (\color{DGLorange} q_1' \color{black} ,q_3';
\color{RoyalBlue} q^{~}_1 \color{black} , q^{~}_4)
\Bigl[\mathcal{S}^\Lambda_{q^{~}_4,q_4'}
\mathcal{G}^\Lambda_{q^{~}_3, q_3'}  
+\mathcal{G}^\Lambda_{q^{~}_4,q_4'}
\mathcal{S}^\Lambda_{q^{~}_3, q_3'}  \Bigr]
\gamma_2^\Lambda (q_4', \color{DGLorange} q_2' \color{black} ;
q^{~}_3, \color{RoyalBlue} q^{~}_2 \color{black}) \; .
 \label{eqgamma2DGL_D}
 \end{align}
\end{subequations}
\end{widetext}
Here the higher order vertices $\gamma_{n \ge 3}$ have already been set to zero.

\subsection{Frequency Parametrisation}
Due to energy conservation, the frequencies in
equations~\eqref{eqgamma1DGL}~and~\eqref{eqgamma2DGL} are not independent: 
\begin{equation}
\begin{array}{l}
 \vspace{2mm}
 \gamma_1(q_1',q_1^{~}) \propto \delta( \omega_{n_1'} - \omega_{n_1^{~}}),\\ 
 \gamma_2 (q_1',q_2';q_1^{~},q_2^{~})  \propto \delta( \omega_{n_1'} + \omega_{n_2'} -
 \omega_{n_1^{~}}- \omega_{n_2^{~}}).
\end{array}
\end{equation}
In the case of the two-particle vertex, this gives a certain freedom to parametrize
its frequency dependence. The natural choice, as will become apparent later on, is to
parametrize it in terms of three bosonic frequencies:
\begin{subequations}
\begin{align}
 \Pi =&\, \omega_{n_1'} + \omega_{n_2'} = \omega_{n_1^{~}} + \omega_{n_2^{~}} \, ,\\
 \Chi =&\, \omega_{n_2'} - \omega_{n_1^{~}} = \omega_{n_2^{~}} - \omega_{n_1'} \, , \\
 \Delta =&\, \omega_{n_1'} - \omega_{n_1^{~}} = \omega_{n_2^{~}} - \omega_{n_2'} \, .
\end{align}
\end{subequations}
Note that due to their definition in terms of purely imaginary frequencies, the bosonic
frequencies are imaginary too. Conversely, the fermionic frequencies can be expressed in terms of the bosonic ones:
\begin{subequations}
 \begin{align}
  \omega_{n_1'} = \frac{1}{2} (\Pi - \Chi + \Delta)\, , & &
  \omega_{n_2'} = \frac{1}{2} (\Pi + \Chi - \Delta) \, , \\
  \omega_{n_1^{~}} = \frac{1}{2} (\Pi - \Chi - \Delta)\, , & &
  \omega_{n_2^{~}} = \frac{1}{2} (\Pi + \Chi + \Delta)\, .
 \end{align}
\end{subequations}

\subsection{Neglecting the Vertex Flow}
For the purpose of treating the inhomogeneous model of equation
\eqref{eqModelGeneral}, we take the quantum number that labels Green's
functions and vertices to denote a composite index of site, spin and
Matsubara frequency, $q_1 = (j_1,\sigma_1,\omega_1)$, etc.  Since the
bare propagators are non-diagonal in the site-index, the number of
independent variables $\gamma_2^\Lambda (q_1',q_2';q_1,q_2)$ generated
by \Eq{eqgamma2DGL} is very large, $\mathcal{O}(N^4 N_{\rm f}^3)$,
where $N_{\rm f}$ is the number of Matsubara-frequencies per
frequency argument kept track of in the numerics.

The simplest way to avoid this complication is to neglect the flow of
the two-particle vertex:
\begin{align}
\label{eqfRG1}
 \frac{d}{d\Lambda} \gamma_2 =0.
\end{align}
This scheme, to be called fRG1, yields a frequency-independent
self-energy, which, for the case of local interaction, is site
diagonal. It is exact to first order in the interaction.

\subsection{Coupled-Ladder Approximation}
For models where the interaction cannot be considered small, we introduced a novel
scheme in Ref.~\onlinecite{Bauer2013}, to be called dynamic fRG in CLA, to
incorporate the effects of vertex flow. In the following, whenever the vertex
flow is included, we treat it using the CLA, thus calling this approximation dfRG2,
to distinguish it from fRG1, and from a static fRG scheme including the vertex flow,
sfRG2, to be introduced later. The dfRG2 scheme exploits the fact that the bare
vertex 
\begin{equation}
\label{eq:bare-vertex}
\begin{array}{l}
v(j_1 \sigma_1,j_2 \sigma_2;j_3 \sigma_3,j_4 \sigma_4) 
=\\ \;  \qquad U_{j_1} 
\delta_{j_1j_2} \delta_{j_3j_4} \delta_{j_1 j_4} \delta_{\sigma_1 \bar{\sigma}_2}
\delta_{\bar \sigma_3 \sigma_4} (\delta_{\sigma_1 \sigma_3}
- \delta_{\sigma_1 \sigma_4}) \; ,
\end{array}
\end{equation}
is purely site-diagonal, and parametrizes the vertex in terms of
$\mathcal{O}(N^2N_{\rm f})$ independent variables.

To this end we consider a simplified version of the vertex flow equation
\eqref{eqgamma2DGL}, where the feedback of the vertex flow is neglected: on the
r.h.s. we replace $\gamma_2^{\Lambda}\to v$. If the feedback of the self-energy
were also neglected, this would be equivalent to calculating the vertex in second order perturbation
theory.  As a consequence, all generated vertex contributions depend on two site indices and a single bosonic
frequency. They have one of the following structures: 
\begin{subequations}
\label{eqStructures}
\begin{equation}
\label{eqStructure_P}
\begin{array}{rl}
P^{\sigma \bar{\sigma}}_{ji} (\Pi) &:= 
\gamma_p^\Lambda 
\left( \color{DGLorange} j\sigma \Pi\! -\! \omega_{n'} , j \bar{\sigma} \omega_{n'}
\color{black} ; 
\color{RoyalBlue} i\sigma \Pi\! -\! \omega_n^{~} , i \bar{\sigma} \omega_n^{~} \color{black} \right)
\\ & \stackrel{\mathcal{O}(v^2)}{\simeq} 
\begin{matrix}
\includegraphics[width=3.5cm]{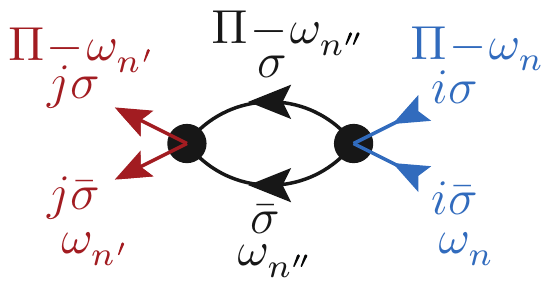}
\end{matrix}\, , 
\end{array}\hspace{-5mm}
\end{equation}
\vspace{-3mm}
\begin{equation}
\label{eqStructure_Pbar}
\begin{array}{rl}
\bar{P}^{\sigma \bar{\sigma}}_{ji} (\Pi ) &:= 
\gamma_p^\Lambda \left( 
\color{DGLorange} j \sigma \Pi\! -\! \omega_{n'}  , j \bar{\sigma} \omega_{n'}
\color{black} ; 
\color{RoyalBlue} i \bar{\sigma} \Pi\! -\! \omega_n^{~} , i \sigma \omega_n^{~} 
\color{black} \right) 
\\ & \stackrel{\mathcal{O}(v^2)}{\simeq}
\begin{matrix}
\includegraphics[width=3.5cm]{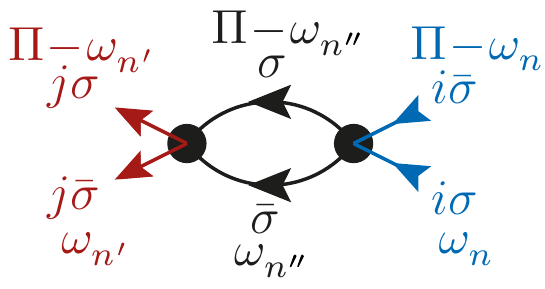}
\end{matrix}\, , 
\end{array}\hspace{-5mm} 
\end{equation}
\vspace{-3mm}
\begin{equation}
\label{eqStructure_X}
\begin{array}{rl}
X^{\sigma \bar{\sigma}}_{ji} ( \Chi ) &:= 
\gamma_x^\Lambda \left( 
\color{DGLorange} j\sigma \Chi\! +\! \omega_{n'} , i \bar{\sigma} \omega_n
\color{black} ; \color{RoyalBlue} i\sigma \Chi\! +\! \omega_n , j \bar{\sigma}
\omega_{n'} \color{black} \right) 
\\ & \stackrel{\mathcal{O}(v^2)}{\simeq}
\begin{matrix}
\includegraphics[width=3.5cm]{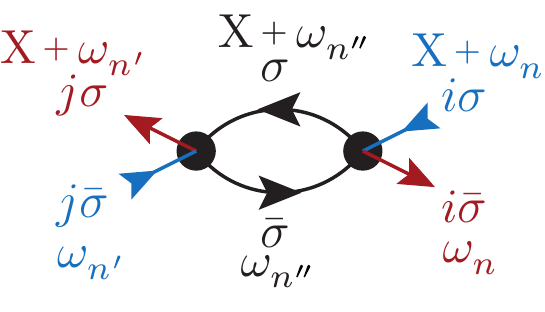}
\end{matrix}\, , 
\end{array}\hspace{-5mm}
\end{equation}
\vspace{-3mm}
\begin{equation}
\label{eqStructure_Xsigma}
\begin{array}{rl}
X^{\sigma \sigma}_{ji} (\Chi ) &:= 
\gamma_x^\Lambda 
\left(\color{DGLorange} j\sigma \Chi\! +\! \omega_{n'} , i \sigma \omega_n
\color{black}; \color{RoyalBlue} i\sigma  \Chi\! +\! \omega_n , j \sigma \omega_{n'} \color{black} \right) 
\\ & \stackrel{\mathcal{O}(v^2)}{\simeq}
\begin{matrix}
\includegraphics[width=3.5cm]{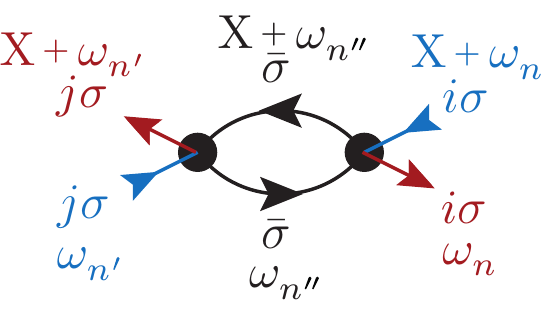}
\end{matrix}\, , 
\end{array}\hspace{-5mm}
\end{equation}
\vspace{-3mm}
\begin{equation}
\label{eqStructure_D}
\begin{array}{rl}
D^{\sigma \sigma}_{ji} (\Delta ) &:= 
\gamma_d^\Lambda 
\left(\color{DGLorange} j\sigma \Delta \! +\! \omega_{n'} , i \sigma \omega_n
\color{black} ; \color{RoyalBlue} j\sigma \omega_{n'} , i \sigma \Delta \! +\!
\omega_n  \color{black} \right) 
\\ & \stackrel{\mathcal{O}(v^2)}{\simeq}
\begin{matrix}
\includegraphics[width=3.5cm]{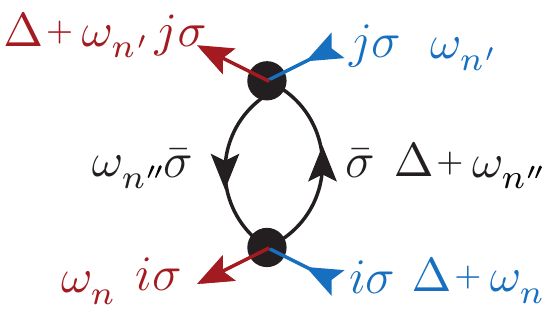}
\end{matrix}\,
\end{array}\hspace{-5mm} 
\end{equation}
\vspace{-3mm}
\begin{equation}
\label{eqStructure_Diso}
\begin{array}{rl}
D^{\sigma \bar{\sigma}}_{ji} (\Delta ) &:= 
\gamma_d^\Lambda 
\left(\color{DGLorange} j \bar{\sigma} \Delta \! +\! \omega_{n'}, i \sigma  \omega_n
\color{black} ; \color{RoyalBlue} j \sigma \omega_{n'} ,i \bar{\sigma} \Delta \! +\!
\omega_n \color{black} \right)
\\ & \stackrel{\mathcal{O}(v^2)}{\simeq}
\begin{matrix}
\includegraphics[width=3.5cm]{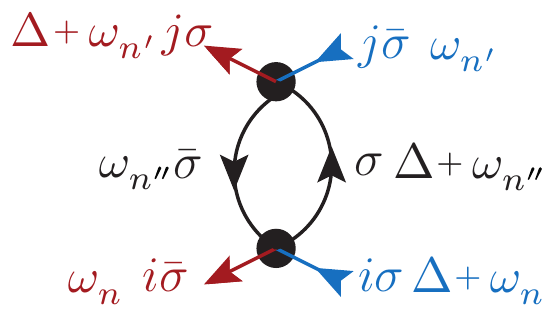}
\end{matrix}\; .
\end{array}\hspace{-5mm} 
\end{equation}
\end{subequations}
These second-order terms do not depend on the frequencies $\omega^{~}_n$ and $\omega_{n'}$. Now note
that no additional terms are generated if we allow for a vertex feedback within the
individual channels in
equations~(\ref{eqgamma2DGL_P},~\ref{eqgamma2DGL_X},~\ref{eqgamma2DGL_D}), i.e.
if we take the flow equation of $\gamma_a (A)$ ($a=p ,x , d$ and correspondingly
 $A=\Pi,\Chi,\Delta$) and replace the feedback of the vertex on the r.h.s. by
 \begin{equation}
 \label{eqRPAScheme}
 \gamma_2(\Pi,\Chi,\Delta) \to v+ \gamma_a(A) \,. 
 \end{equation}
 This scheme is equivalent to solving
 RPA equations for the three individual channels $P$,$X$, and $D$ (see section
 \ref{sec:RPA}), with an additional feedback of the self-energy via
 equation~\eqref{eqfRGgreen}.

Note that if $i\! =\! j$ in equation~\eqref{eqStructures} the terms a and c, b and f as well as d and e
have the same structure w.r.t. their external site and spin indices. As a result, it
is possible to account for an
inter-channel feedback in the vertex flow without generating additional terms if the
feedback is restricted to purely site diagonal terms. As in
Ref.~\onlinecite{Jakobs2010} we avoid frequency mixing by limiting the inter-channel
feedback to the static part of the vertex, i.e. the vertex contributions are evaluated at zero
frequency when fed into other channels. Putting everything together, the
approximation scheme is defined by replacing the vertex on the r.h.s. of the flow
equation $\dot{\gamma}_a^\Lambda$ by 
\eqref{eqgamma2DGL}
\begin{equation}
\label{eqApprScheme}
\gamma_2 \to 
v + \gamma_a (A) + (\gamma_b (0) + \gamma_c (0))
\delta_{j_1j_2}\delta_{j_1'j_2'}\delta_{j_1j_1'}\, ,
\end{equation}
where $a,b,c$ are cyclic permutations of $p,x,d$, and
$A,B,C$ are the corresponding cyclic permutations of the frequencies $\Pi,\Chi,\Delta$.
Since this equation is the central definition of this paper, we explicitly
write it for each of the three channels:
\begin{subequations}
\begin{align}
\begin{array}{rl}
\vspace{1.5mm}
 \hspace{-4.5mm}\dot{\gamma}_p (\Pi) & : \gamma_2(j_1',j_2';j_1,j_2;\Pi,\Chi,\Delta) \\
 \rightarrow & \; v + \gamma_p (\Pi) +
 (\gamma_x (0) + \gamma_d (0))\delta_{j_1j_2}\delta_{j_1'j_2'}\delta_{j_1j_1'} \, ,
 \end{array}
\end{align}
\begin{align}
\begin{array}{rl}
\vspace{1.5mm}
 \hspace{-4.5mm}\dot{\gamma}_x (\Chi) & : \gamma_2(j_1',j_2';j_1,j_2;\Pi,\Chi,\Delta) \\
 \rightarrow & \; v + \gamma_x (\Chi) +
 (\gamma_p (0) + \gamma_d (0))\delta_{j_1j_2}\delta_{j_1'j_2'}\delta_{j_1j_1'}\, ,
 \end{array}
\end{align}
\begin{align}
\begin{array}{rl}
\vspace{1.5mm}
 \hspace{-4.5mm}\dot{\gamma}_d (\Delta) & : \gamma_2(j_1',j_2';j_1,j_2;\Pi,\Chi,\Delta) \\
 \rightarrow & \; v + \gamma_d (\Delta) +
 (\gamma_p (0) + \gamma_x (0))\delta_{j_1j_2}\delta_{j_1'j_2'}\delta_{j_1j_1'} \, .
 \end{array}
\end{align}
\end{subequations}
This scheme generates a self-energy and a vertex which are both exact to
second order in the interaction. To see this we note, that first, the fRG flow
equations without any truncation are exact, and second, in the fRG truncation
\eqref{eqfRGtrunc} and in the CLA \eqref{eqApprScheme} the neglected terms are all of
third or higher order in the interaction.
\vspace{-2mm}
\subsection{Symmetries}
As can readily be checked, these flow equations respect the following symmetry
relations: 
\vspace{-2mm}
\begin{subequations}
\label{eq:vertexsymmetries}
\begin{align}
\label{eq:G-symmetry}
\mathcal{G}^{\sigma \Lambda}_{ij} (\omega_n) &= \mathcal{G}^{\sigma \Lambda}_{ji}
(\omega_n)
= \left[ \mathcal{G}^{\sigma \Lambda}_{ij} (-\omega_n) \right]^* \; , 
\\
\Sigma^{\sigma \Lambda}_{ij} (\omega_n)  &= \Sigma^{\sigma \Lambda}_{ji} (\omega_n)
= \left[ \Sigma^{\sigma \Lambda}_{ij} (- \omega_n) \right]^* \; , 
\end{align}
\end{subequations}
\vspace{-10mm}
\begin{subequations}
\begin{align}
P^{\sigma \bar{\sigma}}_{j i}\! & \! = P^{\bar{\sigma} \sigma}_{j i} = 
P^{\sigma\bar{\sigma}}_{i j}\, ,  \nonumber\\
\bar{P}^{\sigma \bar{\sigma}}_{j i} \! & \! = \bar{P}^{\bar{\sigma} \sigma}_{j i} = 
\bar{P}^{\sigma\bar{\sigma}}_{i j}   \, , \nonumber\\
P^{\sigma \bar{\sigma}}_{j i}\!  & \! = -\bar{P}^{\sigma \bar{\sigma}}_{j i}   \, ,  
\end{align}
\vspace{-10mm}
\begin{align} 
X^{\sigma \sigma'}_{j i} \!  &=  X^{\sigma \sigma'}_{i j} = \Bigl[ X^{\sigma'
\sigma}_{j i} \Bigr]^* \, , \nonumber\\ 
D^{\sigma \sigma'}_{j i} \!  & = D^{\sigma \sigma'}_{i j } = \Bigl[D^{\sigma'
\sigma}_{j i} \Bigr]^* \, , \nonumber\\
X \!  &= - D \, ,
\end{align}
\begin{align}
P^{\sigma \bar{\sigma}}_{j i} (\Pi ) \!  & \!= \left[ P^{\sigma \bar{\sigma}}_{j i}
(-\Pi) \right]^* ,\nonumber\\ 
X^{\sigma \sigma'}_{j i}  (\Chi)  \!  & \!=  
\big[ X^{\sigma \sigma'}_{j i} (-\Chi) \big]^*  \, , \nonumber\\
D^{\sigma \sigma'}_{j i}  (\Delta) \!  & \! =  
\big[ D^{\sigma \sigma'}_{j i} (-\Delta) \big]^*  \, , 
\end{align}
\begin{equation}
X^{\sigma \sigma}, D^{\sigma \sigma} \in \mathbb{R}\; .
\end{equation}
\end{subequations}
As a result only four independent symmetric frequency dependent matrices are left,
which we define as follows:
\begin{align}
\label{eqNonzeroVertex}
P_{ji}^\Lambda (\Pi) &= P^{\sigma \bar{\sigma}}_{ji} (\Pi), \nonumber \\ 
X_{ji}^\Lambda (\Chi) &= X^{\uparrow \downarrow}_{ji} (\Chi) ,\\ 
D^{\sigma \Lambda}_{ji} (\Delta) &= D^{\sigma \sigma}_{ji} (\Delta) \nonumber ,
\end{align}
where the superscript $\Lambda$ signifies a dependence on the flow parameter. At zero
magnetic field the number of independent matrices reduces to three since in this
case $D^\uparrow=D^\downarrow$.

The flow equations for these matrices can be derived starting from
Eqs.~\eqref{eqgamma2DGL}. The replacement
\eqref{eqApprScheme} restricts the internal quantum numbers on the r.h.s.\  of the flow
equation $q_3^{}$, $q_4^{}$, $q_3'$, and $q_4'$ according to
the definitions \eqref{eqStructures}: 
\begin{widetext}
\begin{subequations}
\label{eq:derive-flow-equations-channels-explicit}  
\begin{align}
\label{eq:derive-flow-equations-P-channel}
\dot P^\Lambda_{ji} (\Pi) = &
\dot{\gamma}_p^\Lambda \left( \color{DGLorange} 
j\sigma \Pi\! - \! \omega_{n'} , j \bar{\sigma} \omega_{n'}
\color{black} ; \color{RoyalBlue} 
i\sigma \Pi\! - \! \omega_n , i \bar{\sigma} \omega_n
\color{black} \right) \\ \nonumber
=& T \sum_{kl,n''} 
\Bigl[ 
 \gamma_2^\Lambda \left( \color{DGLorange}
j\sigma \Pi\! - \! \omega_{n'} , j \bar{\sigma} \omega_{n'}
\color{black};
k\sigma \omega_{n''} , k \bar{\sigma}\Pi\! - \!  \omega_{n''}
\right)
\mathcal{S}^{\sigma \Lambda}_{kl}(\omega_{n''}) 
\mathcal{G}^{\bar \sigma \Lambda}_{kl}(\Pi\!-\!  \omega_{n''})
\gamma_2^\Lambda \left( 
l\sigma \omega_{n''} , l \bar{\sigma} \Pi\! - \! \omega_{n''} ;
\color{RoyalBlue} 
i\sigma \Pi\! - \! \omega_n^{} , i \bar{\sigma} \omega_n^{}
\color{black} 
\right)
\Bigr.
\\ \nonumber 
& 
\Bigl. +  
 \gamma_2^\Lambda \left( \color{DGLorange}
j\sigma \Pi\! - \! \omega_{n'} , j \bar{\sigma} \omega_{n'}
\color{black};
k\bar{\sigma} \omega_{n''} , k \sigma\Pi\! - \!  \omega_{n''}
\right)
\mathcal{S}^{\bar{\sigma} \Lambda}_{kl}(\omega_{n''}) 
\mathcal{G}^{\sigma \Lambda}_{kl}(\Pi\!-\!  \omega_{n''} ) 
\gamma_2^\Lambda \left( 
l\bar{\sigma} \omega_{n''} , l \sigma \Pi\! - \! \omega_{n''} ;
\color{RoyalBlue} 
i\sigma \Pi\! - \! \omega_n^{} , i \bar{\sigma} \omega_n^{}
\color{black} 
\right)
\Bigr] \, ,
\end{align}
\begin{align}
\label{eq:derive-flow-equations-X-channel}
\dot X^\Lambda_{ji} (\Chi) =& 
\dot{\gamma}_x^\Lambda \left( \color{DGLorange} 
j\sigma \Chi\! +\! \omega_{n'} , i \bar{\sigma} \omega_n
\color{black} ; \color{RoyalBlue} 
i\sigma \Chi\! +\! \omega_n , j \bar{\sigma} \omega_{n'}
\color{black}
\right) \\ \nonumber
=&  T \sum_{kl,n''}\Bigl[
 \gamma_2^\Lambda \left( \color{DGLorange}
 j\sigma \Chi\! +\! \omega_{n'}
\color{black} ,  
k \bar \sigma \omega_{n''} ;  
k \sigma \Chi\! +\! \omega_{n''} , 
\color{RoyalBlue}
 j \bar \sigma \omega_{n'} \color{black} \right)
\mathcal{S}^{\bar{\sigma} \Lambda}_{kl}(\omega_{n''}) 
\mathcal{G}^{\sigma \Lambda}_{lk}(\Chi \!+\!\omega_{n''} ) 
\gamma_2^\Lambda \left( l\sigma \Chi \!+\!\omega_{n''},  \color{DGLorange} 
i\bar \sigma \omega_n
\color{black} ;  
\color{RoyalBlue} i\sigma \Chi \! +\! \omega_n \color{black}, 
l \bar{\sigma} \omega_{n''} \color{black} \right)\Bigr.
\\ \nonumber
& 
 + \gamma_2^\Lambda \left( \color{DGLorange}
 j\sigma \Chi\! +\! \omega_{n'}
\color{black} ,  
k \bar \sigma \omega_{n''} ;  
k \sigma \Chi\! +\! \omega_{n''} , 
\color{RoyalBlue}
 j \bar \sigma \omega_{n'} \color{black} \right)
\mathcal{G}^{\bar{\sigma} \Lambda}_{kl}( \omega_{n''}) 
\mathcal{S}^{\sigma \Lambda}_{lk}( \Chi \!+\!\omega_{n''} ) 
\gamma_2^\Lambda \left( l\sigma \Chi \!+\!\omega_{n''},  \color{DGLorange} 
i\bar \sigma \omega_n
\color{black} ;  
\color{RoyalBlue} i\sigma \Chi \! +\! \omega_n \color{black}, 
l \bar{\sigma} \omega_{n''} \color{black} \right)\Bigr] \, ,
\end{align}
\begin{align}
\label{eq:derive-flow-equations-D-channel}
\dot D^{\sigma \Lambda}_{ji} (\Delta) = &
\dot{\gamma}_d^\Lambda \left( \color{DGLorange}
j\sigma \Delta\! + \! \omega_{n'}, i \sigma \omega_n
\color{black} ; \color{RoyalBlue} 
j\sigma \omega_{n'} , i \sigma \Delta\! + \! \omega_n
\color{black} \right)  \\
= & -T \sum_{kl,n''} 
\Bigl[ 
\gamma_2^\Lambda \left( \color{DGLorange}
j\sigma \Delta\! + \! \omega_{n'}, \color{black} k \sigma \omega_{n''}
 ; \color{RoyalBlue} 
j\sigma \omega_{n'} , \color{black} k \sigma \Delta\! + \! \omega_n
\right) 
\mathcal{S}^{\sigma \Lambda}_{kl} (\omega_{n''} ) 
\mathcal{G}^{\sigma \Lambda}_{kl} (\Delta \! + \! \omega_{n''} ) 
\gamma_2^\Lambda \left(
l\sigma \Delta\! + \! \omega_{n''}, \color{DGLorange} i \sigma \omega_n
\color{black} ;
l\sigma \omega_{n''} ,\color{RoyalBlue} i \sigma \Delta\! + \! \omega_n
\color{black} \right)
\Bigr.
\nonumber \\ 
& +\gamma_2^\Lambda \left( \color{DGLorange}
j\sigma \Delta\! + \! \omega_{n'}, \color{black} k \sigma \omega_{n''}
 ; \color{RoyalBlue} 
j\sigma \omega_{n'} , \color{black} k \sigma \Delta\! + \! \omega_n
\right) \nonumber 
\mathcal{G}^{\sigma \Lambda}_{kl} (\omega_{n''} ) 
\mathcal{S}^{\sigma \Lambda}_{kl} (\Delta \! + \! \omega_{n''} ) 
\gamma_2^\Lambda \left(
l\sigma \Delta\! + \! \omega_{n''}, \color{DGLorange} i \sigma \omega_n
\color{black} ;
l\sigma \omega_{n''} ,\color{RoyalBlue} i \sigma \Delta\! + \! \omega_n
\color{black} \right)
\Bigr.
\\ 
& +\gamma_2^\Lambda \left( \color{DGLorange}
j\sigma \Delta\! + \! \omega_{n'}, \color{black} j \bar{\sigma} \omega_{n''}
 ; \color{RoyalBlue} 
j\sigma \omega_{n'} , \color{black} j \bar{\sigma} \Delta\! + \! \omega_n
\right) \nonumber 
\mathcal{S}^{\bar{\sigma} \Lambda}_{ji} (\omega_{n''} ) 
\mathcal{G}^{\bar{\sigma} \Lambda}_{ij} (\Delta \! + \! \omega_{n''} ) 
\gamma_2^\Lambda \left(
i\bar{\sigma} \Delta\! + \! \omega_{n''}, \color{DGLorange} i \sigma \omega_n
\color{black} ;
i\bar{\sigma} \omega_{n''} ,\color{RoyalBlue} i \sigma \Delta\! + \! \omega_n
\color{black} \right)
\Bigr.
\\ \nonumber
 &
+\gamma_2^\Lambda \left( \color{DGLorange}
j\sigma \Delta\! + \! \omega_{n'}, \color{black} j \bar{\sigma} \omega_{n''}
 ; \color{RoyalBlue} 
j\sigma \omega_{n'} , \color{black} j \bar{\sigma} \Delta\! + \! \omega_n
\right) 
\mathcal{G}^{\bar{\sigma} \Lambda}_{ji} (\omega_{n''} ) 
\mathcal{S}^{\bar{\sigma} \Lambda}_{ij} (\Delta \! + \! \omega_{n''} ) 
\gamma_2^\Lambda \left(
i\bar{\sigma} \Delta\! + \! \omega_{n''}, \color{DGLorange} i \sigma \omega_n
\color{black} ;
i\bar{\sigma} \omega_{n''} ,\color{RoyalBlue} i \sigma \Delta\! + \! \omega_n
\color{black} \right)
\Bigr] \, .
\end{align}
\end{subequations}
\end{widetext}
As is the case for the diagrams~\eqref{eqStructures}, these equations do not depend on $\omega_n^{~}$ and
$\omega_{n'}$, if the same holds for $\gamma_2$ on the r.h.s.. The latter is of course not the case
without the replacement~\eqref{eqApprScheme}.
The initial conditions are
\begin{align}
&P^{\Lambda_i}=X^{\Lambda_i} = D^{\sigma \Lambda_i} = 0 \; . 
\end{align}

Performing the replacement~\eqref{eqApprScheme}, these equations can be compactly
written in matrix form
\begin{subequations}
\label{eq:matrixODE}
\begin{align}
\label{eq:PODE}
\frac{d}{d\Lambda} P^{\Lambda} (\Pi)  = &
\tilde{P}^{\Lambda} (\Pi) \bubble^{p \Lambda} (\Pi) \tilde{P}^{\Lambda} (\Pi) \; ,\\
\label{eq:XODE}
\frac{d}{d\Lambda} X^{\Lambda} (\Chi)  = &
\tilde{X}^{\Lambda} (\Chi) \bubble^{x \Lambda} (\Chi) \tilde{X}^{\Lambda} (\Chi) \; ,\\
\label{eq:DODE}
\frac{d}{d\Lambda} D^{\sigma \Lambda} (\Delta)  = &-
\tilde{D}^{\sigma \Lambda} (\Delta) \bubble^{\sigma d \Lambda} (\Delta) \tilde{D}^{\sigma \Lambda}
(\Delta) \nonumber \\
 & - I^{\Lambda} \bubble^{\bar{\sigma} d \Lambda} (\Delta)  I^{\Lambda}\; ,
\end{align}
\end{subequations}
where we have introduced the definitions 
\begin{subequations}
\label{eq:crossfeed}
\begin{align}
\label{eq:crossfeed_P}
\tilde{P}^\Lambda_{ji} (\Pi) &=P^\Lambda_{ji}  (\Pi)  + 
\delta_{ji} \left( X^\Lambda_{jj}  (0) + U_j \right),\\
\label{eq:crossfeed_X}
\tilde{X}^\Lambda_{ji} (\Chi) &=
X^\Lambda_{ji} (\Chi) + \delta_{ji} \left( P^\Lambda_{jj} (0) +U_j \right), \\
\label{eq:crossfeed_D}
\tilde{D}^{\sigma \Lambda}_{ji}(\Delta)&=D^{\sigma \Lambda}_{ji} (\Delta) + 
\delta_{ji} X^{\sigma \Lambda}_{jj}(0)
\nonumber \\ &=D^{\sigma \Lambda}_{ji} (\Delta) - \delta_{ji} D^{\sigma \Lambda}_{jj} (0), \\
I^\Lambda_{ji}&=\delta_{ji} \left(P^\Lambda_{jj}(0) + X^\Lambda_{jj} (0)+ U_j\right) \label{eq:Icombination} 
\end{align}
\end{subequations}
which account for the inter-channel feedback contained in equation~\eqref{eqApprScheme}.
$\bubble^p$, $\bubble^x$ and $\bubble^{\sigma d}$ each represent a specific bubble,
i.e.\ a product of two propagators summed over an internal frequency:
\begin{subequations}
\label{eq:bubble}
\begin{align}
\label{eq:Pip}
\bubble^{p \Lambda}_{ji} (\Pi) &=T \sum_{\sigma n} 
 \mathcal{S}^{\sigma \Lambda}_{ji} (\omega_n)
\mathcal{G}^{\bar{\sigma} \Lambda}_{ji} (\Pi\! -\! \omega_n) \; ,\\
\label{eq:Pix}
\bubble^{x \Lambda}_{ji} (\Chi )&=T\sum_{n} 
\Bigl[ \mathcal{S}^{\uparrow \Lambda}_{ji} (\omega_n^{})
\mathcal{G}^{\downarrow \Lambda}_{ij} (\Chi\!+\!\omega_n^{})\nonumber \\
&
\phantom{=T\sum_{n^{}}\Bigl[} + 
 \mathcal{S}^{\downarrow \Lambda}_{ij} (\omega_n^{})
\mathcal{G}^{\uparrow \Lambda}_{ji} (\omega_n^{}\! -\! \Chi)\Bigr],\\
\label{eq:Pid}
\bubble^{\sigma d \Lambda}_{ji} (\Delta )&=T\sum_{n} 
\Bigl[ \mathcal{S}^{\sigma \Lambda}_{ji} (\omega_n^{})
\mathcal{G}^{\sigma \Lambda}_{ij} (\Delta\!+\!\omega_n^{})\nonumber \\
&\phantom{=T\sum_{n}\Bigl[} + 
 \mathcal{S}^{\sigma \Lambda}_{ij} (\omega_n^{})
\mathcal{G}^{\sigma \Lambda}_{ji} (\omega_n^{}\! -\! \Delta)\Bigr],
\end{align}
\end{subequations}

Using the above definitions, the flow equation of the self-energy,
\eqref{eqgamma1DGL}, can be written explicitly as 
\begin{align} \label{eqSelfExplicit} \frac{d}{d\Lambda}
\Sigma_{ji}^{\sigma \Lambda} (\omega_n) = - T & \sum_{n'}  \Bigl[
\bigl(\delta_{ji} U_j + P_{ji} (\omega_n+\omega_{n'})  \nonumber \\ 
& + X_{ji}(\sigma(\omega_n-\omega_{n'})) \bigr) \mathcal{S}^{\bar{\sigma}}_{ji}
(\omega_{n'})
\nonumber \\ \nonumber 
&  - D^\sigma_{ji}(\omega_n-\omega_{n'})  
\mathcal{S}^{\sigma}_{ji} (\omega_{n'}) \\
& +\delta_{ji} \sum_{k} D^\sigma_{jk} (0) \mathcal{S}^{\sigma}_{kk} (\omega_{n'})
\bigr] \;. 
\end{align}

To summarize, dfRG2 is defined by the flow equations
\eqref{eq:matrixODE} and (\ref{eqSelfExplicit}), together with the
definitions \eqref{eqfRGgreen}, \eqref{eqStructures},
\eqref{eqNonzeroVertex}, \eqref{eq:crossfeed} and \eqref{eq:bubble}.

\subsection{Magnetic susceptibility}
In this section we demonstrate how the fRG approach can be used to derive expressions
for linear response theory. We start by defining the magnetic susceptibility
 $\chi_i$ at a given site $i$ as the linear response of the local magnetization
 $m_i$ to a magnetic field $B$:
\begin{eqnarray}
\label{eqchidef}
\chi_i = \left.\partial_B m_i\right|_{B=0} = \frac{1}{2} \left.\partial_B \left( n_i^{\uparrow} -
n_i^{\downarrow}\right)\right|_{B=0},
\end{eqnarray}
where $n_i^{\sigma}$ is the local occupation of site $i$ with
spin $\sigma$.  Using the Matsubara sum representation of the local
density, $n_i^{\sigma} = T \sum_n \mathcal{G}_{ii}^{\sigma}( \omega_n)$, we explicitly calculate the
derivative w.r.t. the magnetic field:
\begin{eqnarray}
\chi_i &=& \frac{T}{2} \sum_{n\sigma} \left. \sigma \partial_B \mathcal{G}_{ii}^{\sigma}(
\omega_n)\right|_{B=0} \notag \\
&=& \frac{T}{2} \sum_{n\sigma} \left. - \sigma \mathcal{G}^{\sigma}( \omega_n)
\partial_B\left(\sigma B/2-\Sigma^{\sigma}( \omega_n)\right)\mathcal{G}^{\sigma}(
\omega_n) \right|_{B=0} \notag \\
&=& -\frac{T}{2} \sum_{nj} \mathcal{G}_{ij}( \omega_n)\mathcal{G}_{ji}( \omega_n) \notag \\
&& +\frac{T}{2}\sum_{nkl\sigma}\sigma \mathcal{G}_{ik}( \omega_n)\partial_B\left. \Sigma_{kl}^{\sigma}(
\omega_n) \right|_{B=0}\mathcal{G}_{li}( \omega_n). 
\end{eqnarray}
Note that the derivative of the self-energy w.r.t. the magnetic field $B$ has the
structure of the fRG flow equation of the self-energy~\eqref{eqgamma1DGL}. So we
perform the derivative by setting $\Lambda\!=\!B$ instead of the $\Lambda$-dependence
defined in equation~\eqref{eqfRGscheme}. The single scale
propagator~\eqref{eqfRGgreen} with $\Lambda\!=\!B$ set to zero then is
\begin{eqnarray}
\mathcal{S}^{\sigma,B=0} =  \mathcal{G} \partial_B \left[\mathcal{G}_0^{\sigma} \right]_{B=0}^{-1}\mathcal{G}=
\frac{\sigma}{2}\mathcal{G}^2.
\end{eqnarray} 

Using this in combination with the flow equation of the self-energy~\eqref{eqgamma1DGL},
\begin{align}
 \partial_B \Sigma_{kl}^{\sigma} (\omega_n) &= \frac{T}{2} \sum_{n' j_1 j_2 j_3
 \sigma'} \sigma'
  \mathcal{G}_{j_1 j_2}^{\sigma'}(\omega_{n'}) 
  \mathcal{G}_{j_2 j_3}^{\sigma'}(\omega_{n'}) \nonumber \\
  & \quad \times \gamma_2 (j_3 \sigma' \omega_{n'}, k \sigma \omega_n ; j_1 \sigma' \omega_{n'} , l \sigma \omega_n  )
   \, ,
\end{align}
one directly arrives at the well known Kubo-formula for the magnetic
susceptibility, which is exact if the self-energy and the vertex are
known exactly. For the coupled-ladder approximation we directly use
the explicit flow equation for the self-energy~\eqref{eqSelfExplicit},
which yields

\begin{eqnarray}
\label{eqchifinal}
\chi_i = &-& \frac{T}{4} \sum_{n,j} \mathcal{G}_{ij}( \omega_n)\mathcal{G}_{ji}( \omega_n) \notag \\
 &+&  \frac{T^2}{2} \sum_{nn'klj}\Bigl( \mathcal{G}_{ik}( \omega_n) \mathcal{G}_{li}( \omega_n)
  \mathcal{G}_{lj}( \omega_{n'})\mathcal{G}_{jk}( \omega_{n'}) \notag \\
 & & \times \left[P_{kl}(\omega_n\!+\!\omega_{n'})\!+\!X_{kl}(\omega_n\!-\!
 \omega_{n'})\!+\!D_{kl}(\omega_n\!-\!\omega_{n'}) \right]\notag \\
 && \; -\,\mathcal{G}_{ik}(\omega_n) \mathcal{G}_{ki}(\omega_n) D_{kl}(0) \mathcal{G}_{lj}(\omega_n)
 \mathcal{G}_{jl}(\omega_n) \Bigr) \,.
\end{eqnarray}

\subsection{Zero-temperature limit}

For the numerical data presented in Sec.~\ref{secfRGresults}, we
  focussed exclusively on the case of zero temperature.  For the fRG
scheme defined by equation \eqref{eqfRGscheme} the limit $T\to 0$ has
to be performed carefully\cite{Karrasch2008}: $\omega_n \to i \omega$
($\omega \in \mathbb{R}$) becomes a continous variable and $\Theta_T$
a sharp step function, with $\Theta (0) = \frac{1}{2} $ and
$\partial_\omega \Theta(\omega) = \delta(\omega)$. For this
combination of $\delta$- and $\Theta$-functions Morris'
lemma\cite{Morris1994} can be appplied, which yields:
\begin{subequations}
\begin{align}
 \mathcal{S}^{\Lambda} (i \omega) &\stackrel{T=0}{=} \delta (|\omega|-\Lambda)
 \widetilde{\mathcal{G}}^{\Lambda} (i \omega), \\
 \widetilde{\mathcal{G}}^{\Lambda} (i \omega) &\; = \; \Big[ \left[
 \mathcal{G}_0 (i\omega )\right]^{-1} - \Sigma^{\Lambda} (i\omega) \Big]^{-1} ,
\\
 \mathcal{S}^{\Lambda}_{i,j} (i \omega_1^{~}) \mathcal{G}^{\Lambda}_{k,l} (i
 \omega_2^{~})
 &\stackrel{T=0}{=} \delta (|\omega_1^{~} |-\Lambda) \Theta ( |\omega_2^{~}
 |-\Lambda) \nonumber\\
  & \qquad \qquad \; \; 
 \widetilde{\mathcal{G}}^{\Lambda}_{i,j} (i \omega_1^{~})
 \widetilde{\mathcal{G}}^{\Lambda}_{k,l} (i \omega_2^{~})\; .
\end{align}
\end{subequations}

\subsection{Static fRG}
A further possible approximation is to completely neglect the
frequency dependence of the vertex. This is done by setting all three
bosonic frequencies $\Pi$, $\Chi$ and $\Delta$ to zero throughout. As
a result, the self-energy is frequency-independent, too. This
approach, called static fRG2 (sfRG2), looses the property of being exact
to second order. It leads to reliable results only for the
zero-frequency Green's function at zero temperature. If knowing the
latter suffices (such as when studying the magnetic field-dependence
at $T=0$), sfRG2 is a very flexible and efficient tool, computationally
cheaper than our full coupled-ladder scheme.

\subsection{Numerical implementation}
\label{secNumericImplementation}
Due to the slow decay of $\mathcal{S}^\Lambda$ for
  $\Lambda\to \infty$, integrating the flow-equation
  \eqref{eqgamma1DGL} of the one-particle vertex $\gamma_1$ from
  $\Lambda=\infty$ to a large but finite value $\Lambda=\Lambda_0$
  yields a finite contribution. 
 For numerical implementations, the initial
  condition thus has to be changed to \cite{Andergassen2004}
\begin{equation}
 \gamma_1^{\Lambda_0}
  ( \color{DGLorange} q_1' \color{black}, \color{RoyalBlue} q^{~}_1
   \color{black}) =
 -\frac{1}{2} \sum_{q} v (q ,\color{DGLorange} q_1' \color{black}; q,
  \color{RoyalBlue} q^{~}_1  \color{black}) \;  .
\end{equation}

All numerically costly steps can be expressed as matrix operations,
for which the optimized toolboxes BLAS and LAPACK can be used.  The
calculation time scales as $\mathcal{O} (N^3)$, due to the occurrence
of matrix inversions \eqref{eqfRGgreen} and matrix products
\eqref{eq:matrixODE}. In the case of sfRG2 there are six matrix
functions, each depending only on $\Lambda$. As a result the
integration is straightforward, and can be done, e.g., by a standard
4th-order Runge-Kutta with adaptive step-size control. We used the
more efficient Dormand-Prince method\cite{Dormand1980}, and mapped the infinite
domain of $\Lambda \in [0,\infty)$ onto a finite domain using the substitution
$\Lambda=\frac{x}{1-x}$ with $x\in [0 ,1)$.  The upper bound for $N$, the maximal
number of sites where $U_j \neq 0$, is mainly set by accessible memory. In the case
of several Gigabytes, $N$ should not exceed $10^4$ to $10^5$.  (We note in passing
that for the one-dimensional Hubbard model [which is a special case of the model
studied below, see Eq.~\eqref{eqModelChain}], $N$-values of that magnitude would not
yet be large enough to reach the Luttiger-liquid regime for the case of small
interactions $U$. The reason is that for the Hubbard model the spectral weight and
the conductance have a non-monotonic dependence on energy: as the energy is
decreased, there is an intermediate regime in which they first increase, before the
power-law decrease characteristic of Luttinger-liquid behavior finally sets in at
very low energy scales, i.e.\ very large system
sizes.\cite{Andergassen2006,Schuricht2013} For small interactions $U \lesssim
0.5\tau$, the latter crossover only becomes accessible for system sizes well
beyond $10^5$ sites (see e.g. Fig.~6 in Ref.~\onlinecite{Andergassen2006}). To be
able to see the low-energy decrease of spectral weight for system sizes of order
$10^5$, interactions would have to be chosen to be as large as $U \simeq 4 \tau$, for
which, however, the CLA can no longer be trusted.)

For dfRG2 all matrices depend additionally on the Matsubara frequency,
which is, in the case of zero temperature, a continuous variable. This
variable has to be discretized in the numerical implementation. A good
and safe choice is a logarithmic discretisation, since analytic
functions have most structure close to their branch cuts, i.e.  small
Matsubara frequencies. Another possible choice, used in
Ref.~\onlinecite{Karrasch2008}, is a geometric mesh. Since an
appropriate discretization consists of at least $100$ frequencies, the
upper bound for $N$ is reduced to $10^3$, for which the run time
already becomes quite large.

For frequency values in between the discrete frequencies on the mesh
the functions have to be interpolated.  Intuitively one might expect
that a nonlinear interpolation, e.g.\ a cubic spline, would lead to
better results. However in our implementations this led to a
self-enhanced oscillatory behavior of the self-energy as function of
frequency, even for a very dense discretization mesh. To avoid such
numerical artifacts, the safest choice is a linear interpolation,
where the density of the discretization is increased until the desired
accuracy is reached.

\subsection{Relation between fRG2 and RPA}
\label{sec:RPA}

In this section we show that in the ladder approximation proposed here, fRG retains the
quality of being closely related to parquet-type equations. This can be seen by
considering a simplified version thereof, in which the coupling between the three
channels is neglected, i.e. using replacement~\eqref{eqRPAScheme} instead of
\eqref{eqApprScheme}, and so is the feedback of the self-energy by replacing
$\tilde{\mathcal{G}}^\Lambda$ by $\mathcal{G}_0$ in equation
\eqref{eq:bubble}. In this case, each of the
three differential equations \eqref{eq:matrixODE} reduces to the generic form,  
\begin{equation}\label{eq:RPA_flow}
 \frac{d}{d \Lambda} \Gamma^{\Lambda} (\nu) = 
\Gamma^{\Lambda} (\nu) \bubble^\Lambda (\nu) \Gamma^{\Lambda} (\nu)
\; , 
\end{equation}
with initial condition $\Gamma^{\Lambda_i} = \mathcal{U} = \delta_{ij} U_j$ 
(with $U_j\geq0$, for present purposes). 
If equation \eqref{eq:RPA_flow} converges, its solution is given by
\begin{subequations}
\label{eq:RPA}
\begin{eqnarray}
\Gamma  (\nu)
&  =  &  \mathcal{U} \left[ \mathbb{I} + \bubble (\nu) \mathcal{U}  \right]^{-1} \; ,
\end{eqnarray}
with
\begin{eqnarray}
\bubble (\nu) & = & \int^{\infty}_{0} 
d \Lambda \bubble^{\Lambda} (\nu) \; .
\label{eq:bubble_gen}
\end{eqnarray}
\end{subequations}

Now note that \Eq{eq:RPA} is also obtained if each channel is
separately treated in the Random Phase Approximation (RPA).
Consequently, the full fRG2 scheme (either dynamic or static), described by \Eqs{eq:matrixODE},
amounts to a simultaneous treatment of all RPA-channels with
inter-channel coupling via \eqref{eq:crossfeed}, and a feedback of
Hartree type diagrams via \eqref{eqfRGgreen}.

\section{fRG results} \label{secfRGresults}

In this section we will discuss some properties of the results obtained with the
fRG-equations stated in section \ref{secfRG}, for the case of a QPC-geometry. We will
compare the results for the linear response conductance for the three approximation
schemes and discuss the spin-susceptibility within dfRG2.

\subsection{Model for a QPC}
We note that Eq.~\eqref{eqModelGeneral} applies to systems of arbitrary spatial
dimensions. However, in this work we only present and discuss results for QPCs, thus
restricting the model to one dimension. The lowest one-dimensional subband of the QPC
is modeled by an inhomogeneous tight-binding chain, with onsite interactions:
\begin{align}
 \hspace{-3.5mm} H\!=\! \sum_{j\sigma} [E_{j}^{\sigma} n_{j\sigma} \! - \! \tau
 (d^\dagger_{j\sigma} d^{}_{j+1\sigma} + {\rm h.c.})] + \sum_j U_j
 n_{j\uparrow}n_{j\downarrow} ,
 \label{eqModelChain}
\end{align}
with $E_{j}^{\sigma} = V_j + 2 \tau - \frac{\sigma B}{2} $ where $B$ is a Zeeman-field. For low kinetic energies this
tight-binding model is a good approximation for a continuum model with mass
$\frac{m}{\hbar^2}=\frac{1}{2\tau a^2} $ (where $\hbar$ is Plank's constant ) and potential
$V_j=V(x=ja)$\cite{Forsythe1960}, provided that the lattice spacing $a$ is much smaller than the
length scales on which the potential changes. In order to keep computational time
small, the model should always be chosen in such a way that the number of sites $N$
where $V_j$ or $U_j$ are nonzero is as small as possible. In other words: The
inhomogeneity should be incorporated within as few sites as possible, without loss
of adiabaticity.

We model the QPC as a smooth one-dimensional potential barrier which is purely
parabolic around its maximum at $x=0$:
\begin{align}
V(x)=V_{g} + \mu - \frac{m}{2\hbar^2}\Omega_x^2 x^2,
\end{align}
or in discrete version:
\begin{align}
V_j=V_{g} + \mu - \frac{\Omega_x^2}{4\tau} j^2 \qquad (|j| < j_c).
\label{eqPot}
\end{align}
Here, $j_c$ defines the range of pure parabolicity, $\mu$ is the chemical potential
and $\Omega_x$ is the relevant energy scale for the QPC,\cite{Buttiker1990} which we
define such that it has the dimension of an energy (not frequency). The condition
that $a$ has to be much smaller then the length scales on which the potential
changes, implies the condition $\Omega_x \! \ll \! \tau$.  $V_{g}$ is the
gate-voltage, which controls the height of the potential. For $|j| > j_c$ the
potential is smoothly connected to homogenous semi-infinite noninteracting leads. The
potential can be considered as purely parabolic regarding its low-energy transport
properties if $j_c\gg \sqrt{{\tau}/{\Omega_x} }$. In the following we use
$\mu=0.5\tau$, $\Omega_x=0.04\tau$, $j_c=\sqrt{2\tau \mu}/\Omega_x$ and $N=81$. These
values optimize the conditions on $\Omega_x$, $j_c$ and the smoothness of the
potential on the one hand and the smallness of the number of sites $N$ on the other
hand.  Typical experimental values for GaAs QPCs are $\Omega_x=1{\rm meV}$ and
$m=0.067 m_e$, where $m_e$ is the electron mass. The latter fixes the hopping to
$\tau=25 {\rm meV}$ and thus the length unit to $a=\sqrt{{\hbar^2}/{2\tau m}}\simeq 5
{\rm nm}$. These values should give a rough estimate for comparison with experiment,
however, in the following we will use the system of measurement defined by $\tau$ and
$a$, without referring to SI units.

\subsection{Model Properties}
\label{secResutsModel}
\begin{figure}
 ~\hspace{-3mm}
\includegraphics[width = 89mm]{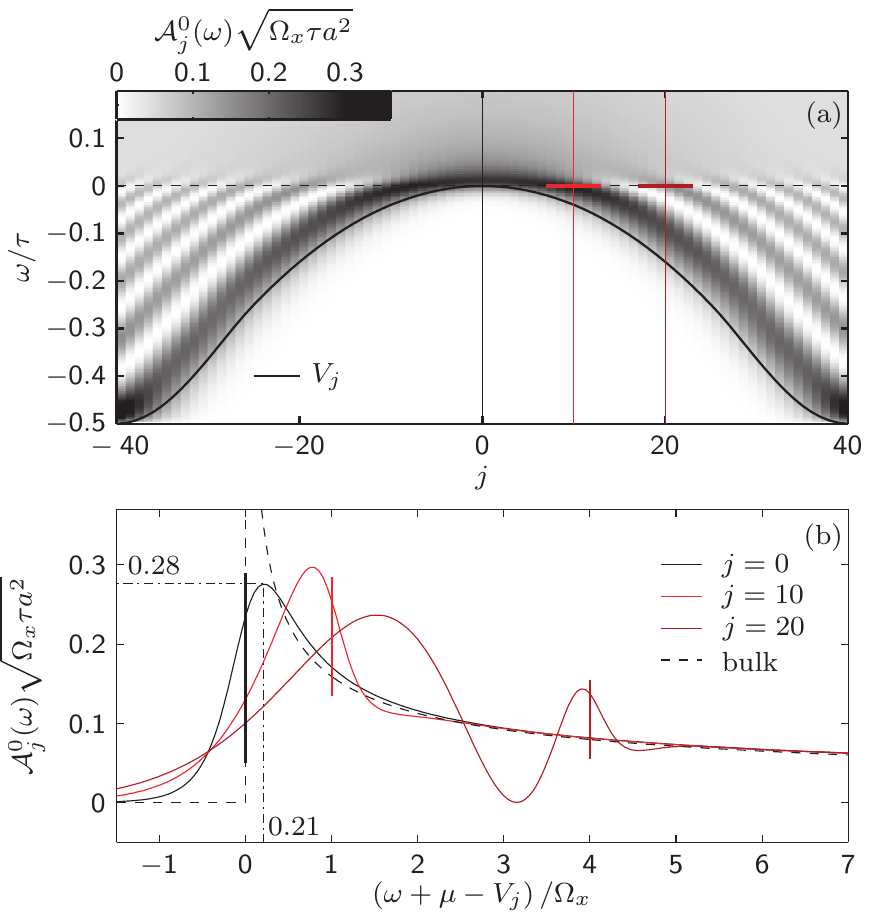}\\
\caption{\small (a) Local Density of States $\mathcal{A}_j(\omega)$ (color
scale) for the noninteracting, $U_j=0$, Hamiltonian Eq.~\eqref{eqModelChain} with
potential~\eqref{eqPot} at $V_g\!=\!0$ (thick black line). (b) Local Density of States
$\mathcal{A}_j (\omega)$ as a function of $(\omega-V_j)/\Omega_x$ for a
homogenous tight binding chain ($V_j=0$, grey line) and for the
potential~\eqref{eqPot} at fixed site $j=0$ (blue), $j=10$ (green) and $j=20$ (red),
indicated in (a) by vertical lines with corresponding colors.}
\label{figModProp}
\end{figure}

Having defined the model we first discuss its noninteracting ($U\! =\! 0$) properties. Figure
\ref{figModProp} shows the local density of states (LDOS) 
\begin{align}
\mathcal{A}_j(\omega) = -\frac{1}{a \pi} {\rm Im} \mathcal{G}_{jj} (\omega + i 0^+)
\end{align}
both in a greyscale plot as a function of site index and frequency (a) and
at several fixed sites as a function of frequency (b). Note that just above the
potential [black line in Fig.~\ref{figModProp}(a)] the LDOS is enhanced [dark region
in Fig.~\ref{figModProp}(a)]. This property originates from the fact that the density
of states (DOS) of a one-dimensional system shows a divergence at zero velocity:
indeed the DOS for the homogenous version [$V_j\! =\! 0$, i.e. $V_g\! =\! \mu\! =\!
\Omega_x\! =\! 0$
in Eq.~\eqref{eqPot}] of our model [black
dashed line in Fig.~\ref{figModProp}(b)] reads
\begin{align}
 \mathcal{A}(\omega) =\frac{1}{\pi a\sqrt{\omega(4\tau - \omega)}} \stackrel{\omega \ll
 \tau}{\approx}
 \frac{1}{2 \pi a \sqrt{\tau \omega}}  \propto \frac{1}{v^{\rm clas}} ,
\end{align}
where $v^{\rm clas}$ is the classical velocity of the electron.
Quantum-mechanically, this divergence is smeared out by the inhomogeneity ($V_j\!
\neq \! 0$) of a potential. Following Ref.~\onlinecite{Bauer2013}, we call this smeared van
Hove singularity in the LDOS that follows the potential a ``van Hove ridge''. In the
case of a parabolic barrier with curvature given by $\Omega_x$ \eqref{eqPot} the
maximum of the LDOS is at an energy of $\mathcal{O}(\Omega_x)$ bigger than $V_j$ and
has a height of $\mathcal{O}(\sqrt{\tau \Omega_x})$ [see dashed-dotted line in
Fig.~\ref{figModProp}(b)]. For energies below the potential maximum electrons get
reflected. This leads to standing waves, altering the LDOS by oscillations around its
bulk value [white striped area in Fig.~\ref{figModProp} (a) and oscillations in dark
red line Fig.~\ref{figModProp} (b)].
 \begin{figure*}
\includegraphics[width =183mm]{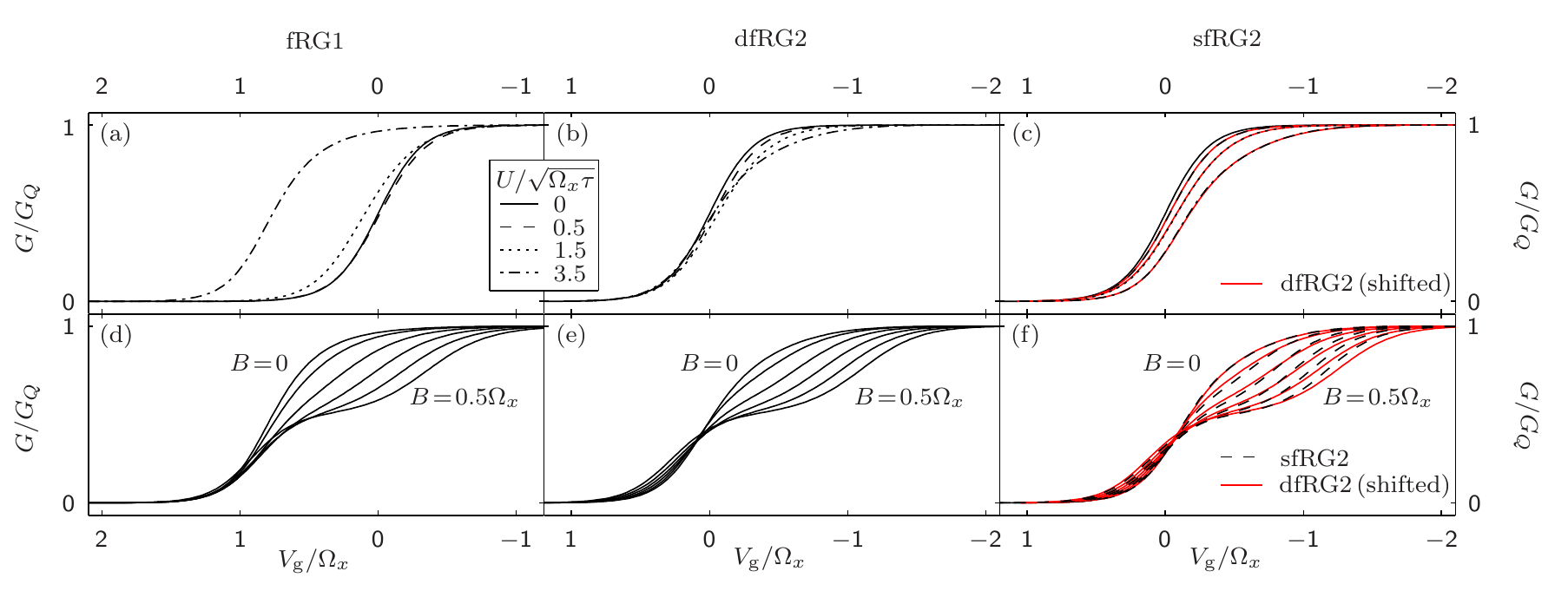} \caption{\small
(a) to (c) Conductance $G$, as a function of gate-voltage, $V_g$, at zero magnetic field,
$B\!=\!0$, for different values of interaction $U$. (d) to (f) Conductance $G$ at fixed
interaction strength $U=3.5\sqrt{\Omega_x}$, for six equidistant magnetic fields $B$, between
$0$ and $0.5\Omega_x$. Conductance is calculated using fRG1 (a, d),
dfRG2 (b, c) or sfRG2 black lines in (c, f). Red lines in (c, f) show
dfRG2 data repeated from (b, e) with a $U$-dependent shift, $\Delta V_g$, in
$V_g$-direction ($\Delta V_g = 0$, $0.001$, $0.02$ and $0.15\Omega_x$ for $U=0$,
$0.5$, $1,5$ and $3.5 \sqrt{\Omega_x \tau}$ respectively)} 
\label{figfRG1vsdfRG}
\end{figure*}

\subsection{Conductance of a QPC}
\label{secCond}
Having discussed the properties of the noninteracting model, we
continue with the fRG-results at finite interaction. For this we first
define the spatial dependence of the interaction $U_j$, which, for the
one-dimensional model is an effective one-dimensional interaction
resulting from integrating out two space dimensions. Its strength
depends on the geometry, and is larger if the spatial confinement
perpendicular to the one-dimensional system is smaller. We assume that
this confinement is independent of the position in the transport
direction in the center of the QPC, with $U_{j=0}\!=\!U$. This is a
fair assumption for a saddle point approximation of the
two-dimensional QPC potential. For $|j|\to N'=\frac{N-1}{2}$, $U_j$
drops smoothly to zero, describing the adiabatic coupling to the
two-dimensional electron system, represented by the semi-infinite
tight binding chain.

In Ref.~\onlinecite{Bauer2013} we showed that the 0.7-anomaly is
caused by the van Hove ridge in the LDOS discussed above. Its apex
crosses the chemical potential when the QPC is tuned into the sub-open
regime, i.e. the regime where the conductance takes values
$0.5G_Q<G<0.9 G_Q$. This high LDOS at the chemical potential enhances
effect of interactions by two main mechanisms: first, the effective
Hartree barrier depends nonlinearly on gate-voltage and magnetic
field, causing an enhanced elastic backscattering; and second, due to
the high LDOS inelastic backscattering is enhanced once a phase space
is opened up by a finite temperature or source-drain voltage. Both
effects reduce the conductance in the sub-open regime, causing the
0.7-anomaly. Since interactions are enhanced by the LDOS, the relevant
dimensionless interaction strength is $U_j\mathcal{A}_j(\mu)a$, which
scales like $U/\sqrt{\Omega_x\tau}$ in the sub-open regime.

In this paper we will concentrate on examining how the reliability of
the method depends on the interaction, without explaining the physical
mechanism underlying the 0.7-anomaly in detail (for the latter we
refer to Ref.~\onlinecite{Bauer2013}). For the model
Eq.~\eqref{eqModelChain} no reliable results are availible from other
methods to which we could have compared our own. Instead, we here
compare the results of the different fRG-schemes fRG1, sfRG2 and
dfRG2. These schemes differ in the prefactor of the perturbative
expansion of terms in order $U^2$ and higher. If these terms are
important the three approximation schemes will deviate from each
other.  Hence, the qualitative and quantitative reliability can be
deduced from the qualitative and quantitative deviations between these
schemes.

The first observable we discuss is the linear response conductance at zero
teperature\cite{Datta1997}:
\begin{equation}
\label{eqcond}
 G=\frac{e^2}{h} \sum_\sigma \left| 2\pi \rho^{\sigma}(i0^+)
 \mathcal{G}_{-N'N'}^{\sigma} (i0^+) \right|^2 \, ,
\end{equation}
where $\rho(\omega)$ is the density of states at the boundary of a
semi-infinite tight-binding chain, representing the two-dimensional
leads (for a derivation of the boundary Green's function, see appendix
Sec.~\ref{secProjection}).

Particularly interesting in studying the 0.7-anomaly in QPCs is the
shape of the conductance trace as a function of applied gate voltage
in the region where its value (in units of $G_Q$) changes from zero to
one, and how this shape changes with external parameters, such as
applied magnetic field. First of all, we emphasize the good
qualitative agreement of all three approximation schemes with each
other as well as with experimental results, compare
Figs.~\ref{figfRG1vsdfRG} (d), (e) and (f) with
Ref.~\onlinecite{Thomas1996,Cronenwett2002} (A direct comparison of
dfRG2 with experiment is given in Ref.~\onlinecite{Bauer2013}). This
confirms that the method qualitatively captures the physical mechanism
with respect to the conductance at zero temperature very well.

For a more quantitative analysis we first consider the position of the
conductance step, say $V_{\rm po}$; even though the actual position of
the step is of minor interest experimentally, it gives information
about how accurate Hartree-type correlations are treated.
Figs.~\ref{figfRG1vsdfRG} (a), (b), and (c) show the conductance at
$B=0$ for increasing values of interaction $U$ for fRG1, dfRG2, and
sfRG2 respectively. While for dfRG2 and for sfRG2 $V_{\rm po}$
decreases with interaction, its behavior for fRG1 is non-monotonic:
$V_{\rm po}$ decreases slightly at small values of interaction, and
increases strongly at larger values of interaction. Hence the
conductance at large interaction is higher than the bare, $U=0$,
value. This behavior is unphysical: whenever the density is nonzero,
an increase in $U$ should cause an increase in the effective barrier
height due to coulomb repulsion, and hence a decrease in the
conductance. This artifact is significantly reduced by the vertex-flow
of both dfRG2 and sfRG2. For the latter, interactions suppress the
conductance more strongly than for the former. Due to these deviations
between the three schemes,  we cannot make a quantitative statement
about the exact position of the conductance step $V_{\rm po}$.

The deviations just discussed make quantitative comparisons between these
methods (or with others, such as RPA) difficult if interactions are large. The reason
for the difficulty is that the $V_g$-position of the conductance step depends
sensitively on the precise way in which Hartree-type correlations are treated and
hence differ for each of the above schemes. Hence it would not be meaningful to
compare their predictions for physical quantities calculated \textit{at a given
value} of $V_{g}$; instead, it is only meaningful to compare the shape of their
evolution with $V_{g}$.  Actually, the same is true for physical quantities that are
dominated by Fock-type correlations, since internal propagators have to be dressed by
Hartree diagrams. Doing this is crucial for inhomogeneous systems such as ours, since
an inhomogeneous density causes these Hartree contributions to have a significant
dependence on position and gate voltage.  In the fRG approach the feedback of the
self-energy Eq.~\eqref{eqfRGgreen} always guarantees that internal lines are dressed
in a self-consistent way.

Let us now compare the shapes of the $V_g$-dependent conductance
curves for dfRG2 and sfRG2. To this end, replotted the dfRG2 data from
Fig.~\ref{figfRG1vsdfRG} (b) in Fig.~\ref{figfRG1vsdfRG} (c), with a
$U$-dependent shift $\Delta V_g$ in $V_{g}$-direction (red lines). It
can be seen from comparison with sfRG2 data, that the shapes of the
conductance curves are almost identical.

Next we analyze the shape of the conductance step at finite
interaction, and how it develops with magnetic field. The effect of an
increasing magnetic field is qualitatively similar for the three
approximation schemes [see Figs.~\ref{figfRG1vsdfRG} (d), (e) and
(f)]: the conductance step develops into a spin resolved double step,
in an asymmetric way: while the curves hardly change for $V_g$ values
where $G<0.5 G_Q$, they are strongly suppressed in the sub-open
regime, where the LDOS is large. For fRG1 the $V_g$-range where the
lowest magnetic field, $B=0.1\Omega_x$, significantly reduces the
conductance w.r.t.\ the conductance at $B\!  =\! 0$, is larger than
for dfRG2 and sfRG2. This is related to the fact, that the
magnetoconductance, the change of conductance with magnetic field,
within fRG1 is negative even for $V_g$ values where conductance is
close to zero [this effect is too small to be visible in
Fig.~\ref{figfRG1vsdfRG} (d)].  Since this is not the case for dfRG2
and sfRG2 it is not possible to make a reliable statement about the
sign of the magnetoconductance in the tunnel regime. Again we
reproduced the dfRG2 data from Fig.~\ref{figfRG1vsdfRG} (e) in
Fig.~\ref{figfRG1vsdfRG} (f) with a shift $\Delta V_g$ in $V_g$ (red
line) in order to compare their shape with the sfRG2 data (black
dashed line). The effect of the magnetic field on the conductance
within sfRG is slightly larger for small fields and slightly smaller
for large fields, than for the dfRG2 results.  Based on the facts,
that, first, the deviations between dfRG2 and sfRG2 are small and,
second, sfRG2 is computationally much cheaper than dfRG2, we conclude
that for preliminary studies, or when scanning a large parameter
space, one should favor sfRG2 whenever it is sufficient to know the
static part of vertex functions.

\subsection{Susceptiblity}

\begin{figure}
\includegraphics[width = 8.9cm]{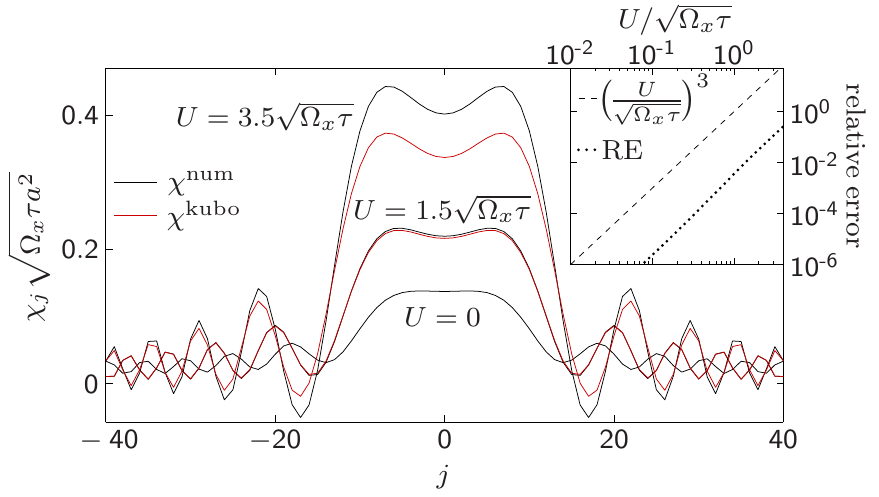}
\caption{\small Local spin-susceptibility $\chi_j$,
  Eq.~\eqref{eqchidef},as a function of site index, for the QPC
  potential Eq.~\eqref{eqPot}, at $V_g\!=\!-\Omega_x/4$, calculated
  using dfRG2 via the the numerical derivative of the local
  magnetisation, $\chi^{\rm num}$ (black lines), as well as via the
  Kubo formula \eqref{eqchifinal}, $\chi^{\rm kubo}$ (red lines), for
  three different values of interaction. Inset: relative error ${\rm
    RE}$~\eqref{eqRE} (dots), as a function of interaction $U$, on a
  log-log scale. The error scales as $U^3$ (compare dashed line),
  since dfRG2 is exact to second order in the interaction $U$.}
\label{figSus}
\end{figure}
As explained in reference~\onlinecite{Bauer2013}, the 0.7-anomaly is
related to an enhanced spin susceptibility in the sub-open regime of
the QPC. For this quantity an estimate of the error is available
within the dfRG2 approximation scheme.  We note that the
spin-susceptibility defined in equation~\eqref{eqchidef} can be
calculated in two ways: by numerical differentiation of the
magnetization for a small magnetic field, $\chi^{\rm num}$, or via
equation~\eqref{eqchifinal}, $\chi^{\rm kubo}$. Like the conductance,
the value of $\chi$ is not known exactly.  Thus we argue here as in
the previous section. $\chi^{\rm num}$ and $\chi^{\rm kubo}$ are both
exact to second order in the interaction, as can easily be checked,
but they differ in terms that are of order $U^3$ and higher. If the
difference of $\chi^{\rm num}$ and $\chi^{\rm kubo}$ is significant,
the higher order terms are non-negligible, and the results cannot be
trusted.

In reference~\onlinecite{Bauer2013} we showed that $\chi^{\rm num}_j$
is enhanced due to the inhomogeneity of the QPC potential and in
addition amplified by interactions. It has a strong $V_g$-dependence,
and is maximal when the QPC is tuned into the sub-open regime. In this
regime, at $V_g=-\Omega_x/4$, we compare $\chi^{\rm num}$
(Fig.~\ref{figSus} black lines) with $\chi^{\rm kubo}$
(Fig.~\ref{figSus} red lines) for different values of interaction. For
intermediate values of interaction $U=1.5\sqrt{\Omega_x\tau}$ both
results are essentially identical, while for a larger value of
interaction $U=3.5\sqrt{\Omega_x\tau}$ deviations are clearly visible,
however still not too large. The qualitative features that the
susceptibility strongly increases with interaction, and that it is
enhanced in the center of the QPC, emphasized in
Ref.~\onlinecite{Bauer2013}, are confirmed by both results.
Furthermore they coincide in their spatial structure, i.e. two maxima
in the center and a decaying oscillating behavior. This spatial
structure is mainly given by the LDOS at the chemical potential (see
Sec.~ \ref{secResutsModel}) and enhanced by interactions.

For a better quantification we define the relative error:
\begin{equation}
\label{eqRE}
{\rm RE}=
2\frac{\sum_j \left| \chi_{j}^{\rm kubo} -\chi_{j}^{\rm num}\right|}
{\sum_j\left|\chi_{j}^{\rm kubo}+\chi_{j}^{\rm num}\right| }\, . 
\end{equation}
This error is shown on a log-log scale in the inset of
Fig.~\ref{figSus} (dots).  The relative error scales like $U^3$, since
dfRG2, and thus $\chi^{\rm num}$ and $\chi^{\rm kubo}$ are exact to
second order in $U$. For the larger value of interaction,
$U=3.5\sqrt{\tau\Omega_x}$, the relative error of about $18\%$ is
quite significant and thus the value of $\chi$ is quantitatively not
reliable. The reason for this is that the dimensionless interaction
strength $U_j\mathcal{A}_j(\mu)a\approx 3.5 \cdot 0.3 \approx 1$ is
already close to one. Nevertheless the error is still dominated by the
third order term, implying that it is controlled.

Finally we note that the spin-susceptibility in the RPA approximation
\begin{equation}\label{eqSusRPA}
 \chi_i^{\rm RPA} =\sum_j \left[ \bubble^d (0) \left[1 + \mathcal{U} \bubble^d (0)
 \right]^{-1} \right]_{ij} \; . 
\end{equation}
diverges at an interaction strength for which fRG is still well-behaved.  For
example, if bare internal propagators are used to calculate $\bubble^d$, $\chi_i^{\rm
RPA}(V_g)$ turns out to diverge at $U\simeq 3.3 \sqrt{\Omega_x \tau}$. Moreover, the
value of $\chi^{\rm RPA}$ and thus also the $U$-value for which it diverges depends on how
one chooses to treat interactions for internal propagators of $\bubble^d$.  However
RPA itself gives no recipe how to do this. In contrast, the fRG approach gives a
systematic framework for computing the two-particle vertex, the self-energy, and
their feedback into each other, in a way that moderates competing instabilities in an
unbiased way (as mentioned in section III). 

\section{Conclusion and Outlook}

We have derived a fRG based approximation scheme, called the
coupled-ladder approximation (CLA), for spin-full fermionic tight
binding models with a local interaction and an arbitrary
potential. The main applications are systems with a significant
spatial dependence, in particular, models where the electron density
significantly changes with the position in real space.

The CLA is formulated within the context of third-order-truncated fRG schemes,
in which the three particle vertex is set to zero, while the flow of the two particle
vertex is fully incorporated. The CLA retains two of the main properties of
third-order-truncated fRG: it is exact to second order, and sums up diagrams of
the RPA in all channels.  Since the CLA is based on a perturbative argument, it
is reliable only if interactions are not too large.

We analyzed in detail the reliability of this approach for a one-dimensional tight
binding model with a parabolic potential barrier representing a QPC. For this we
compared results for the conductance and the spin susceptibility calculated using
different approaches within the fRG for different interactions up to $U\! =\!
3.5\sqrt{\Omega_x \tau}$. We identified the magnetic field dependence of the
conductance and the enhanced susceptibility related to the 0.7
anomaly,\cite{Bauer2013} as robust properties of the model.  

Finally, let us comment briefly on the prospects of using the CLA approach presented
here to obtain finite-temperature results.  While the formulas for the local density
$n$ and the local susceptibility $\chi$, Eq.~\eqref{eqchifinal}, are valid for
arbitrary temperature $T$, the conductance formula Eq.~\eqref{eqcond} holds only for
the case of zero temperature. The generalization of this formula to finite
temperature\cite{Oguri2001} involves an analytic continuation to the real axis for
both self-energy and vertex w.r.t.\ their frequency arguments.
However, performing such an analytic continuation for numerical data is a
mathematically ill-defined problem and turns out to be especially difficult for
matrix-valued functions. 

One possibility to avoid this complication is to formulate our CLA
scheme on the Keldysh contour, in which case there are several
different possibilities for introducing the fRG flow parameter
\cite{Jakobs2009a}. (For a fRG treatment of the Single Impurity
Anderson Model see Ref.~\onlinecite{Jakobs2010}). When using Keldysh
fRG to treat equilibrium properties, the number of independent
correlators can be reduced by exploiting the Kubo-Martin-Schwinger
conditions.\cite{Jakobs2009} Moreover, Keldysh fRG in principle also
allows non-equilibrium properties to be calculated.  The actual
implementation of Keldysh fRG for our model will be nontrivial,
though, in particular since numerical integrations along the real
axis, where Green's functions can have poles, can be
challenging. Another problem at finite temperature is the
  violation of particle conservation due to the fRG truncation
  \eqref{eqfRGtrunc},\cite{enssThesis}. The extent of this violation
  might be reduced by by implementing the modified vertex flow
  suggested by Katanin.~\cite{Katanin2004} We believe that it
would be worth pursuing work in these directions.

\begin{acknowledgments}
  We thank Sabine Andergassen, Severin Jakobs, Volker Meden, and
  Herbert Sch\"oller for very helpful  discussions. We acknowledge support
  from the DFG via SFB-631, SFB-TR12, De730/4-3, and the Cluster of
  Excellence \emph{Nanosystems Initiative Munich}.
\end{acknowledgments}

\appendix
\section{Projection method}
\label{secProjection}
The propagator in the fRG flow Eqs.~\eqref{eqgamma1DGL} and
\eqref{eqgamma2DGL}, in general, lives on an infinite-dimensional
chain. However, since the interacting region has finite extent, we
only need to evaluate it on an $N$-dimensional subspace.  Furthermore,
for the evaluation of Eq.~(\ref{eqchifinal}) we need to calculate the
sum over all site indices $j$, including the infinite number of sites
in the leads.  To this end we perform a standard projection
procedure\cite{Karrasch2006,Taylor1972}.  With this method the
influence of the leads on the propagator and their contribution to the
sum can be calculated analytically if the diagonalization of the leads
is known analytically.  Therefore we define projection operators $C$
and $L$, with $C^2 =C$, $L^2=L$ and $L+C=\mathbb{1}$ which divide the
Hilbert space into the part that describes the leads, $L$, and the
finite dimensional part that describes the central region where
interaction is non-zero, $C$. Furthermore, we define for a given quadratic
Hamiltonian $H$ (for an interacting system $H$ is replaced by $H_0\!
+\!\Sigma$), $H_c=CHC$, $H_c=CHC$, $H_{cl}=CHL$, $H_{lc}=LHC$,
$\omega_l=\omega L$ and $\omega_c=\omega C$ and write the Hamiltonian
in the basis defined by the projections: \begin{eqnarray} H =
  \left( \begin{array}{ccc}
      H_{c} & H_{cl}  \\
      H_{lc} & H_{l} \end{array} \right).
 \end{eqnarray}
Consequently the Greens function in the same basis reads
\begin{eqnarray}
\mathcal{G} &=& 
\begin{pmatrix}
\omega_c-H_{c} & -H_{cl}  \\
 -H_{lc} &  \omega_l-H_{l} 
\end{pmatrix}^{-1}
=\begin{pmatrix}
\mathcal{G}_c & \mathcal{G}_{cl}  \\
\mathcal{G}_{lc} &  \mathcal{G}_l   .
\end{pmatrix}
\end{eqnarray} 
with
\begin{subequations}
\begin{align}
\mathcal{G}_c&=\frac{1}{\omega_c-H_c -H_{cl}g_l H_{lc}}\, , 
& g_l = \frac{1}{\omega_l - H_l} \, , \\
\mathcal{G}_l &= \frac{1}{\omega_l-H_l -H_{lc}g_cH_{cl}} \, ,
& g_c = \frac{1}{\omega_c - H_c} \, ,\\ 
\mathcal{G}_{cl} &=\mathcal{G}_c H_{cl} g_l= g_c H_{cl}\mathcal{G}_l
\label{eqGreencl} \, ,\\
\mathcal{G}_{lc} &= g_l H_{lc} \mathcal{G}_c = \mathcal{G}_l H_{lc} g_c \, .
\label{eqGreenlc} 
\end{align}
\end{subequations}
In the following we will only use $\mathcal{G}_l$ and $g_c$ as well as
$G_{cl}$ and $G_{lc}$ expressed in terms $\mathcal{G}_l$ and $g_c$, so
we use the shorthands $\mathcal{G}=\mathcal{G}_l$ and $g=g_l$ (whether
$\mathcal{G}$ lives on the infinite-dimensional Hilbert space, or on
the subspace of the central contact, will be clear from its site
indices).

For the case of the infinite tight-binding chain defined by
Eq.~\eqref{eqModelChain}, the central region extends from site $\! -\!
N'$ to site $N'$, with $N'=\frac{N-1}{2}$, and the coupling to the
leads can be expressed as:
\begin{subequations}
\label{eqHamiltoncl}
\begin{eqnarray}
H_{cl} &=& -\tau
\left(d^\dagger_{-N'}d^{~}_{-N'-1} + d^\dagger_{N'} d^{~}_{N'+1}\right)\\
H_{lc} &=& H_{cl}^{\dagger} \; .
\end{eqnarray}
\end{subequations}
Consequently
\begin{eqnarray}
H_{cl}gH_{lc} &=& 
\tau^2 \left(d^\dagger_{-N'}d^{~}_{-N'-1} + d^\dagger_{N'}d^{~}_{N'+1} \right) 
\notag \\ 
&& \times g
\left(d^\dagger_{-N'-1}d^{~}_{-N'} + d^\dagger_{N'+1} d^{~}_{N'} \right)\notag \\
&=& 
\tau^2 b \left( n_{-N'} + n_{N'}\right),
\end{eqnarray}
where $b \!= \!g_{N'+1,N'+1}$ is the boundary Greens function of a
half-infinite tight binding chain.  Transforming into $k$-space and
using the boundary condition $\langle d^\dagger_{N'} d_k\rangle=0$ we
get $\langle d^\dagger_{N'+1}d_k\rangle \propto \sin^2(k)$. Together
with the dispersion $\varepsilon_k = -\mu-2\tau\cos(k)$ and the proper
normalisation, this yields for ${\rm Im} (\omega_n)>0$:
\begin{eqnarray}
b(\omega_n) &=& \frac{1}{\pi} \int_{-\pi}^{\pi} dk
\frac{\sin^2(k)}{\omega_n + \mu +2\tau \cos(k)} \notag \\
&=& \frac{1}{2\tau^2}\left[\omega_n\! +\! \mu -i~
\sqrt{4\tau^2-(\omega_n+\mu)^2}\right] . \qquad \end{eqnarray}
The square root is defined to have a positive real part, and
$b(-\omega_n)=b^*(\omega_n)$. (For the spin-dependent boundary Green's function
at finite magnetic field, $\mu$ has to be replaced by $\mu+\sigma B/2$).

Next we calculate the infinite sum in Eq.~(\ref{eqchifinal}). We split the sum
into three parts and take $k$ and $l$ to be site indices in the central region.
\begin{align}
& \sum_{j=-\infty}^{\infty} \mathcal{G}_{kj}\mathcal{G}_{jl} =
\left(\sum_{j=-\infty}^{-N'-1}+\sum_{j=-N'}^{N'}+\sum_{j=N'+1}^{\infty}\right)
\mathcal{G}_{jk}\mathcal{G}_{jl} = \notag \\ 
&=  \sum_{j=-N'}^{N'} \mathcal{G}_{jk}\mathcal{G}_{jl} 
 + \tau^2 \mathcal{G}_{k,-N'}h^L \mathcal{G}_{-N',l}
 +  \tau^2 \mathcal{G}_{k,N'} h^R \mathcal{G}_{N',l} , 
\end{align}
with 
\begin{subequations}
\begin{align}
h^L&=\sum_{j=-\infty}^{-N'-1}g_{-\!N'\!-\!1,j} g_{j,\!-\!N'\!-\!1} , \\
h^R&=\sum_{j=N'+1}^{\infty}g_{N'+1,j}g_{j,N'+1} , 
\end{align}
\end{subequations}
where we made use of Eqs.~\eqref{eqGreencl}, \eqref{eqGreenlc} and
\eqref{eqHamiltoncl}.

Finally we note that the last two terms are identical and given by
\begin{eqnarray}
h(\omega_n)\! &\!=\!&\! h^L(\omega_n) = h^R(\omega_n) = \left[ g^2(i
\omega_n) \right]_{N'+1,N'+1} = \notag \\ 
&\!=\!& \!  \frac{1}{\pi} \int_{-\pi}^{\pi} dk
\frac{\sin(k)^2}{\left(\omega_n+\mu+2\tau \cos(k)\right)^2} = \notag \\ &\!=\!&\!
\frac{1}{2\tau^2}\! \left(\frac{\omega_n+\mu}{\omega_n
+\mu-2\tau}\sqrt{\frac{\omega_n +\mu-2\tau}{
\omega_n+\mu+2\tau}}-1\right)\! .\;\; \qquad
\end{eqnarray}

\end{document}